\documentstyle[12pt,amssymb,graphics]{article}
\input epsf

\def\rulerheight{0.5pt}

\def\U1{$U(1)$}
\def\SU5{$SU(5)$}
\def\SO10{$SO(10)$}
\def\422{$SU(4)\otimes SU(2)_L \otimes SU(2)_R$}

\def\diag.{\hbox{diag.}}

\def\M_U{\hbox{$M_U$}\ }
\def\M_P{\hbox{$M_P$}\ }

\def\MSSM+N{\hbox{MSSM+$\nu$}}
\def\etal{{\it et al.}}

\def\ibid{{\it ibid.}}

\def\olbar#1{\overline{#1}}

\newcommand\beq{\begin{equation}}
\newcommand\eeq{\end{equation}}
\newcommand\bea{\begin{eqnarray}}
\newcommand\eea{\end{eqnarray}}
\newcommand\beann{\begin{eqnarray*}}
\newcommand\eeann{\end{eqnarray*}}
\newcommand\ba{\begin{array}}
\newcommand\ea{\end{array}}

\begin{document}
\baselineskip 24pt
\newcommand{\sheptitle}
{Global Analysis of a Supersymmetric Pati-Salam Model }

\newcommand{\shepauthor}
{T. Bla\v{z}ek$^*$, S. F. King and J. K. Parry}

\newcommand{\shepaddress}
{Department of Physics and Astronomy, University of Southampton \\
        Southampton, SO17 1BJ, U.K}

%%%%%%%%%%%%%%%%%%%%%%%%%%%%%%%%%%%%%%%%%%%%%%%%%%%%%%%%%%%%%%%%%%%%%%%%%%%%%%%
%                                  ABSTRACT
%%%%%%%%%%%%%%%%%%%%%%%%%%%%%%%%%%%%%%%%%%%%%%%%%%%%%%%%%%%%%%%%%%%%%%%%%%%%%%%
\newcommand{\shepabstract}
{We perform a complete global phenomenological analysis of
a realistic string-inspired model
 based on the supersymmetric Pati-Salam $SU(4)\times SU(2)_L \times SU(2)_R$ 
 gauge group supplemented by a $U(1)$ family symmetry,
and present predictions for all observables including
muon $g-2$, $\tau \rightarrow \mu \gamma$,
and the CHOOZ angle. 
Our analysis demonstrates the compatibility of such a model with
all laboratory data including charged fermion masses and
mixing angles, LMA MSW and atmospheric neutrino masses and 
mixing angles, and $b \rightarrow s \gamma$,
allowing for small deviations from third family 
Yukawa unification. We show that in such
models the squark and slepton masses may be rather light
compared to similar models with exact Yukawa unification.}

\begin{titlepage}

\begin{flushright}
hep-ph/0303192
\end{flushright}
\begin{center}
{\large{\bf \sheptitle}}
\\ \shepauthor \\ \mbox{} \\ {\it \shepaddress} \\ 
{\bf Abstract} \bigskip \end{center} \setcounter{page}{0}
\shepabstract
\begin{flushleft}
\today
\end{flushleft}

\vskip 0.1in
\noindent
$^*${\footnotesize On leave of absence from 
the Dept. of Theoretical Physics, Comenius Univ., Bratislava, Slovakia}

\end{titlepage}

\newpage
\section{Introduction\label{intro}}
The origin of fermion masses and mixing angles represents a challenge
faced by theorists for a long time. In the post-SuperKamiokande era 
this puzzle has become more intriguing than ever before.
SuperKamiokande evidence for atmospheric neutrino oscillations 
\cite{SKamiokandeColl} has taught us that neutrino masses are non-zero 
and furthermore that the 23 mixing angle is almost maximal.  More recently 
SNO \cite{Ahmad:2002jz} and KamLAND \cite{kamland_exp} experimemts have
confirmed the matter enhanced Large Mixing Angle(LMA) solution to 
the solar neutrino problem \cite{GoPe}. 
In this work we assume that the smallness of
neutrino masses can be explained by the see-saw
mechanism involving very heavy right-handed neutrino states,
and that the see-saw mechanism is implemented using 
single right-handed neutrino dominance \cite{SRHND} which can
explain in a natural way the coexistence of large neutrino mixing
angles with a mass hierarchy. It then becomes a flavour problem
to fit together the neutrino mass puzzle with the pieces provided by
the long-known pattern of quark and charged lepton masses.

The flavour problem cannot be fully addressed without unification.
However, unification has its own challenges.
These include the unification of gauge couplings and third family
Yukawa couplings and the introduction of supersymmetry. 
It is well known that supersymmetry facilitates gauge coupling unification, 
stabilises the hierarchy between the high energy scale and
the weak scale, and allows a radiative mechanism of electroweak
symmetry breaking. Within the natural framework of supersymmetric
unification, the larger high energy gauge group in turn
increases the predictive power of the theory in the flavour sector,
for example by leading to group theoretical mass relations
between quark and lepton masses of the same family.
Relations between quarks and leptons of different families 
require an additional family symmetry, however.
In this way it becomes possible to address both the flavour problem
and the unification problem, within a single framework.
Having defined the framework, it is by no means guaranteed that
models exist which satisfy all the phenomenological constraints
provided by current data, and comply with all the theoretical
requirements such as successful electroweak symmetry breaking,
and approximate gauge and Yukawa unification, while reproducing
the known observables. It is therefore
important to know that at least some models exist which satisfy all
the constraints, as an existence proof that such a proceedure 
can be implemented consistently.

In this paper we shall study a particular example of a complete
supersymmetric unified model of flavour, based on the Pati-Salam 
$SU(4)\times SU(2)_L\times SU(2)_R$ gauge group \cite{PaSa}
extended by an additional $U(1)$ family symmetry.
Accepting minimality as a model building principle
this group has the following nice features: 
it establishes the third family Yukawa unification, 
places the right-handed neutrinos into non-trivial multiplets
and does not introduce unwanted exotic states in the multiplets 
containing the Standard Model fermions and two Higgs doublets
required by its SUSY extension. The Pati-Salam group can emerge from 
a simple gauge group like $SO(10)$ or $E(6)$. However, from a 
string theory perspective, it is not necessary in order to achieve
unification that there should be a unified field theory based
on a simple group. A partially unified gauge group can equally well
emerge directly from string theory, and in the case of the Pati-Salam
gauge group this possibility has been explored extensively
both in the case of weakly coupled fermionic string theories
\cite{AnLe} and in the case of type I strings with D-branes
\cite{ShTy}.

Although models based on the Pati-Salam gauge group have been 
extensively examined, there is currently no complete up to date
phenomenological study of this model in the literature.
For instance \cite{KiOl_LFV} investigated constraints 
from  Lepton Flavour Violation(LFV) in a Pati-Salam model with
small neutrino mixing angles. Subsequently a  Pati-Salam model
was proposed \cite{KiOl_nu}, using single right-handed neutrino
dominance \cite{SRHND} to achieve naturally large neutrino mixing
angles, but the question of LFV was not readdressed
and it was later shown \cite{BlKi_1} that the
$\mu\to e\gamma$ branching ratio is too large.
Moreover, only the negative $\mu$ parameter
was considered in \cite{KiOl_nu,KiOl_yu} which is currently disfavoured by
the muon $g-2$. In other works such as 
\cite{PaSa_Gomez1} and \cite{PaSa_Gomez2}
the neutrino sector is absent all together.
The complete lepton sector is studied in great detail in a global
analysis in \cite{BlKi_1}, but the quark mass matrices used
\cite{KiOl_nu} were obtained for the opposite sign of $\mu$, 
and the analysis gives imperfect fits for 
% the observables in the quark sector like 
the branching ratio $BR(b\to s\gamma)$ or $b$ quark
mass $m_b$ which both get potentially significant contributions from
SUSY loops proportional to $\mu$. To summarise, 
a completely phenomenologically acceptable supersymmetric
Pati-Salam model does not currently exist 
in the literature. This illustrates the broader point that while 
many models exist in the literature, it is less common for 
the analysis of any such model to be complete.

In this paper, then,
we shall construct a ``4221'' model, following the approach of
\cite{KiOl_nu}, and demonstrate its phenomenological viability. 
The model has approximate 
third family Yukawa unification perturbed by higher order 
terms and assumes non-universal soft Higgs masses.
To demonstrate the viability of such a model, we perform a top down global
analysis of the parameter space carried out on 24 observables. 
In the leptonic sector the observables include 
the muon $g-2$ and solar and atmospheric neutrino data.
A complete list of observables and their $\sigma$ values, which are used 
to calculate the $\chi^2$ function can be found in Table \ref{t:observables}.
%  
%  
%  Our global analysis includes a fit to the present muon $g-2$ discrepancy 
%  between Standard Model prediction \cite{muon_SM} 
%  and experiment \cite{muon}.  
%  This discrepancy presently lies at $34\times 10^{-10}$, a $3\sigma$ 
%  effect.\footnote{The value of this discrepancy will depend upon whether 
%  one favours the Standard Model prediction determined from $e^{+}e^{-}$ 
%  data or from $\tau$ decay data \cite{g-2_tau}.  Here we have favoured 
%  the $e^{+}e^{-}$ results, but have carried out a study of the effect 
%  of a lower $g-2$ discrepancy covering the value predicted by $\tau$ 
%  decay data.}  A fit to quark sector mass and mixing, and the branching 
%  ratio for $b\rightarrow s\gamma$ has also been included in our global 
%  analysis.  
%  
%  
%  
%  
%  
In the analysis we ensure that the upper limits on the branching ratio 
for the lepton flavour violating processes $\tau \rightarrow \mu \gamma$, 
$\mu \rightarrow e \gamma$ and $\tau \rightarrow e \gamma$ are not exceeded
as well as the limit on the 13 neutrino mixing angle.
In addition to this an experimental lower bound on each sparticle mass was 
imposed. In particular,
the most constraining are: the LEP limits on the charged SUSY masses
($m_{\tilde{\chi}^\pm},m_{\tilde{\tau}}>105$GeV), the CDF limit
on the mass of the $CP$ odd Higgs state 
($m_{A^0}>105$-$110\,$GeV, valid for $\tan\beta\approx 50$) \cite{Tevatron}, 
and the requirement that the lightest SUSY particle should be neutral.

Having constructed the model and demonstrated its phenomenological
viability, we then discuss the following three aspects of the model in 
more detail:

\begin{itemize}

\item The first such aspect, as first pointed out in \cite{BlKi_1},
is lepton flavour violation arising from the large 23 neutrino mixing 
through a neutrino Yukawa texture of the form 
\beq
Y^{\nu}_{LR}
\sim 
\left( \begin{array}{ccc}
0 & 0 & 0    \\
0 & 0 & 1    \\
0 & 0 & 1   
\end{array}
\right).
\eeq
Due to large $\tan\beta$ additional features emerge when
studying correlations among observables like BR($b\rightarrow s \gamma$), 
BR($\tau\rightarrow \mu \gamma$) and muon $g-2$. Most notably,
two distinct minima are found with similar $\chi^2$ values for the
best fits. These conclusions
are new since study \cite{BlKi_1} did not investigate a complete
model and all other previous works did not involve global
analysis.

\item  The preference for positive $\mu$, given by the sign of 
the muon $g-2$ discrepancy,
implies positive gluino corrections to $m_b$ 
thus leading to difficulties in obtaining $t-b-\tau$ Yukawa unification.  
Hence a second focus of the present work is to study the required 
deviation from third family Yukawa unification in the best fits.
\footnote{This was also
recently studied from a somewhat different point of view in
\cite{PaSa_Gomez1}.}

\item Thirdly we focus on the effects of future experimental advances, 
in the form of direct Higgs searches, a lepton flavour violating 
$\tau \rightarrow \mu \gamma$ measurement and a refinement of the 
muon $g-2$ discepancy, upon our global fits, indicating how further
experimental progress in these areas will constrain the parameter
space of the model.

\end{itemize}

The remainder of the paper is layed out as follows.
Section 2 briefly reviews our construction
of a string-inspired Pati-Salam model. Section \ref{procedure} contains 
a brief description of the numerical technique used in the
analysis.  A discussion of our main results can be found in 
section \ref{res_dis}, with concluding remarks in section \ref{conc}.

\section{The Model\label{model}}

The model considered in this paper 
is based on the Pati-Salam gauge group \cite{PaSa},
supplemented by a $U(1)$ family symmetry,
\begin{equation}
\mbox{SU(4)}\times \mbox{SU(2)}_L \times \mbox{SU(2)}_R  \times U(1)
\label{422}
\end{equation}
The left-handed quarks and leptons are accommodated in the following
representations,
\begin{equation}
{F^i}^{\alpha a}=(4,2,1)=
\left(\begin{array}{cccc}
u^R & u^B & u^G & \nu \\ d^R & d^B & d^G & e^-
\end{array} \right)^i
\end{equation}
\begin{equation}
{\olbar{F}}_{x \alpha}^i=(\bar{4},1,\bar{2})=
\left(\begin{array}{cccc}
\bar{d}^R & \bar{d}^B & \bar{d}^G & e^+  \\
\bar{u}^R & \bar{u}^B & \bar{u}^G & \bar{\nu} 
\end{array} \right)^i
\end{equation}
where $\alpha=1\ldots 4$ is an SU(4) index, $a,x=1,2$ are
SU(2)$_{L,R}$ indices, and $i=1\ldots 3$ is a family index.  The Higgs
fields are contained in the following representations,
\begin{equation}
h_{a}^x=(1,\bar{2},2)=
\left(\begin{array}{cc}
  {h_2}^+ & {h_1}^0 \\ {h_2}^0 & {h_1}^- \\
\end{array} \right) \label{h}
\end{equation}
(where $h_1$ and $h_2$ are the low energy Higgs superfields associated
with the MSSM.) The two heavy Higgs representations are \cite{AnLe}
\begin{equation}
{H}^{\alpha b}=(4,1,2)=
\left(\begin{array}{cccc}
u_H^R & u_H^B & u_H^G & \nu_H \\ d_H^R & d_H^B & d_H^G & e_H^-
\end{array} \right) \label{H}
\end{equation}
and
\begin{equation}
{\olbar{H}}_{\alpha x}=(\bar{4},1,\bar{2})=
\left(\begin{array}{cccc}
\bar{d}_H^R & \bar{d}_H^B & \bar{d}_H^G & e_H^+ \\
\bar{u}_H^R & \bar{u}_H^B & \bar{u}_H^G & \bar{\nu}_H 
\end{array} \right). \label{barH}
\end{equation}

The Higgs fields are assumed to develop VEVs,
\begin{equation}
<H>\equiv<\nu_H>\sim M_{GUT}, \ \ <\olbar{H}>\equiv<\bar{\nu}_H>\sim M_{GUT}
\label{HVEV} 
\end{equation}
leading to the symmetry breaking at $M_{GUT}$
\begin{equation}
\mbox{SU(4)}\otimes \mbox{SU(2)}_L \otimes \mbox{SU(2)}_R
\longrightarrow
\mbox{SU(3)}_C \otimes \mbox{SU(2)}_L \otimes \mbox{U(1)}_Y
\label{422to321} 
\end{equation}
in the usual notation.  Under the symmetry breaking in
Eq.\ref{422to321}, the Higgs field $h$ in Eq.\ref{h} splits into two
Higgs doublets $h_1$, $h_2$ whose neutral components subsequently
develop weak scale VEVs,
\begin{equation}
<h_1^0>=v_1, \ \ <h_2^0>=v_2 \label{vevs}
\end{equation}
with $\tan \beta \equiv v_2/v_1$.

%in the superfield multiplets
%\begin{equation}
%{F^i_{L,R}}=
%\left(\begin{array}{cccc} 
%{u}  & {u} & {u} & {\nu} \\  
%{d} & {d} & {d} & {e^-}     
%\end{array} \right)_{L,R}^i     
%\end{equation}
%We also have two sets of Higgs, the first $h$ contains the two MSSM Higgs doublets
%and transforms as $h\sim (1,2,2)$
%\begin{equation}
%h=
%\left(\begin{array}{cc}
%{h_1}^0 & {h_2}^+ \\   
%{h_1}^- & {h_2}^0      
%\end{array} \right) 
%\end{equation}
%and the second set, $H,\olbar{H}$, transform as 
%$H\sim (4,1,2)$, $\olbar{H}\sim (\bar{4},1,2)$ 
%and develop VEVs which break the Pati-Salam group.
%Finally there are the Pati-Salam singlets, $\theta, \bar{\theta}$,
%which develop VEVs that break the $U(1)$ family symmetry.
%\begin{equation}
%{H},\olbar{H} =
%\left(\begin{array}{cccc}
%{u_H} & { u_H} & { u_H} & { \nu_H} \\
%{ d_H} & { d_H} & { d_H} & { e_H^-}   
%\end{array} \right),\cdots
%\end{equation}
%We assume for convenience that all symmetry breaking
%scales are at the GUT scale, $<H>=<\olbar{H}>=<{ \nu_H}>\sim M_{GUT}$ and $<\theta>=<\bar{\theta}>\sim M_{GUT}$.
%\begin{equation}
%<H>=<\olbar{H}>=<{ \nu_H}>\sim M_{GUT} \sim 10^{16}GeV
%\end{equation}
%\begin{equation}
%<\theta>=<\bar{\theta}>\sim M_{GUT} \sim 10^{16}GeV
%\end{equation}

To construct the quark and lepton mass matrices
we make use of non-renormalisable operators \cite{Ops} of the form:
\begin{eqnarray}
&\left.i\right)& \hspace{1cm} (F^i \olbar{F}^j )h\left(\frac{H\olbar{H}}{M^2}\right)^n
\left(\frac{\theta}{M}\right)^{p_{ij}}\label{Yuk_op}\\
&\left.ii\right)& \hspace{1cm} (\olbar{F}^i\olbar{F}^j )\left(\frac{HH}{M^2}\right)
\left(\frac{H\olbar{H}}{M^2}\right) ^m
\left(\frac{\theta}{M}\right) ^{q_{ij}}.\label{Maj_op}
\end{eqnarray}
The $\theta$ fields are Pati-Salam singlets 
which carry $U(1)$ family charge and develop VEVs which break the 
$U(1)$ family symmetry.
They are required to be present in the operators above to balance the
charge of the invariant operators.
After the $H$ and $\theta$ fields acquire VEVs, they generate a hierarchy in 
$\left.i\right)$ effective Yukawa couplings and $\left.ii\right)$ Majorana masses.  These 
operators are assumed to originate from additional interactions at the scale $M>M_{GUT}$.  The 
value of the powers $p_{ij}$ and $q_{ij}$ are determined by the assignment of $U(1)$ charges,
with $X_{\theta}=-1$ then $p_{ij}=(X_{F^{i}}+X_{\olbar{F}^{j}}+X_{h})$ and 
$q_{ij}=(X_{\olbar{F}^{i}}+X_{\olbar{F}^{j}}+X_{h})$.  

The contribution to the third family Yukawa coupling is assumed to be only from the renormalisable operator with $n=p=0$ leading to Yukawa unification. 
The contribution of an operator, with a given power $n$, to the matrices 
$Y_{f=u,d,\nu,e},\,M_{RR}$ is determined by the 
relevant Clebsch factors coming from the gauge contractions within
that operator.  A list of Clebsch factors for all $n=1$ operators can
be found in the appendix of \cite{KiOl_nu}.  These Clebsch factors give
zeros for some matrices and not for others, hence a choice of
operators can be made such that a large 23 entry can be given to
$Y_{\nu}$ and not $Y_{u,d,e}$.  We shall write, 
\beq
\delta = \frac{<H><\olbar{H}>}{M^2}=0.22, \ \ \epsilon =
\frac{<\theta>}{M^2}=0.22, 
\label{eq:de}
\eeq
then we can identify $\delta$ with mass
splitting within generations and $\epsilon$ with splitting between
generations.  

Our choice of $U(1)$ charges are as in \cite{KiOl_nu}, 
and can be summarised as 
$X_{F^{i}}~=~\left(1,0,0\right)$, 
$X_{\olbar{F}^{i}}~=~\left(4,2,0\right)$, 
$X_{h}=0$, $X_{H}=0$ and $X_{\olbar{H}}=0$.  
This fixes the powers of $\epsilon$ in each entry of the Yukawa
matrix, but does not specify the complete operator.
The Yukawa couplings are specified by the choice of operators,
\begin{equation}
Y_{f}(M_{GUT}) =\left(\matrix{
({a_{11}\cal O}^R+a_{11}^{\prime\prime}{\cal O}^{\prime\prime V})\epsilon^5  &
(a_{12}{\cal O}^I+a_{12}^{\prime}{\cal O}^{\prime F})\epsilon^3 &
(a_{13}^{\prime}{\cal O}^{\prime c})\epsilon \cr
(a_{21}{\cal O}^G)\epsilon^4 &
(a_{22}{\cal O}^W+a_{22}^{\prime}{\cal O}^{\prime S})\epsilon^2 &
(a_{23}{\cal O}^I+a_{23}^{\prime}{\cal O}^{\prime W}) \cr
(a_{31}{\cal O}^R)\epsilon^4 &
(a_{32}{\cal O}^M+a_{32}^{\prime}{\cal O}^{\prime K})\epsilon^2 &
a_{33} }\right)
\label{SRNDmRLopMtr}
\end{equation}
The operators are defined in \cite{KiOl_nu},
although the selection of operators here is different from that paper.
The notation is such that
${\cal O}$, ${\cal O}^{\prime}$ and ${\cal O}^{\prime\prime}$
are $n=1$, $n=2$ 
and (highly small) $n=3$ operators respectively
where $n$ refers to the powers of $(H\olbar{H})$, thus~
\footnote{The $n=3$ operators can, to a very good approximation, 
be neglected. Their inclusion here serves only to fill the 11 entries
of the $Y_{u,\nu}$ Yukawa matrices, thereby ensuring (for example) 
that the up quark is given a very small mass.}
\beq
{\cal O}\sim (H\olbar{H})\sim \delta, \ \ 
{\cal O}^{\prime}\sim (H\olbar{H})^2\sim \delta^2, \ \ 
{\cal O}^{\prime\prime}\sim (H\olbar{H})^3\sim \delta^3.\label{OO'O''}
\eeq
The order unity coefficients $a_{ij},a_{ij}^{\prime},a_{ij}^{\prime\prime}$
multiply the operators ${\cal O},{\cal O}^{\prime},{\cal O}^{\prime\prime}$
in the $ij$ position.
The Majorana operators are assumed to arise from an $m=0$ operator
in the 33 position and $m=1$ operators elsewhere.
The operator choice in Eq.\ref{SRNDmRLopMtr} leads to the
quark and lepton mass matrices in Table \ref{GUT}. For example the Clebsch
coeficients from the leading ${\mathcal{O}}^{W}$ operator in the 22 positon
gives the ratio $0:1:3$ in the $Y_{U,D,E}$ matrices. This ratio along with subleading corrections provides the correct $m_{c}:m_{s}:m_{\mu}$ ratio.
\begin{table}
\bea
Y_u(M_{GUT}) & = & \left(\matrix{
                      \sqrt{2}\:a_{11}^{\prime\prime}\delta^3\epsilon^5 & 
                        \sqrt{2}\:a_{12}^{\prime}\delta^2\epsilon^3 & 
{\displaystyle \frac{2}{\sqrt{5}}}\:a_{13}^{\prime}\delta^2\epsilon   \cr
       \mbox{\rule[-8mm]{0mm}{17mm}}
                                                       0     & 
{\displaystyle \sqrt{\frac{343}{670}}}\:a_{22}^{\prime}\delta^2\epsilon^2 &
                                                       0     \cr
                                                       0     & 
{\displaystyle \frac{8}{5}}\:a_{32}^{\prime}\delta^2\epsilon^2 &
                                                       r_ta_{33}  \cr}
                              \right)
\nonumber \\
\noalign{\bigskip}
Y_d(M_{GUT}) & = & \left(\matrix{
             {\displaystyle \frac{8}{5}}\:a_{11}\delta  \epsilon^5 &
          -\sqrt{2}\:a_{12}^{\prime}\delta^2\epsilon^3 & 
{\displaystyle \frac{4}{\sqrt{5}}}\:a_{13}^{\prime}\delta^2\epsilon   \cr
       \mbox{\rule[-8mm]{0mm}{17mm}}
{\displaystyle \frac{2}{\sqrt{5}}}\:a_{21}\delta  \epsilon^4 &
{\displaystyle \sqrt{\frac{   2}{  5}}}\:a_{22}\delta  \epsilon^2   +
 {\displaystyle \sqrt{\frac{1372}{670}}}\:a_{22}^{\prime}\delta^2\epsilon^2 &
 {\displaystyle \sqrt{\frac{   2}{  5}}}\:a_{23}^{\prime}\delta^2           \cr
 {\displaystyle \frac{8}{5}}\:a_{31}\delta  \epsilon^5 &
     \sqrt{2}\:a_{32}\delta  \epsilon^2 &
                                                        r_ba_{33}   \cr}
                                  \right)
\nonumber \\
\noalign{\bigskip}
Y_e(M_{GUT}) & = & \left(\matrix{
  {\displaystyle \frac{6}{5}}\:a_{11}\delta  \epsilon^5 &
                                                        0    & 
                                                        0    \cr
       \mbox{\rule[-8mm]{0mm}{17mm}}
{\displaystyle \frac{4}{\sqrt{5}}}\:a_{21}\delta  \epsilon^4 &
-3\,{\displaystyle \sqrt{\frac{  2}{  5}}}\:a_{22}\delta  \epsilon^2   +
{\displaystyle \sqrt{\frac{772}{670}}}\:a_{22}^{\prime}\delta^2\epsilon^2 &
-3\,{\displaystyle \sqrt{\frac{  2}{  5}}}\:a_{23}^{\prime}\delta^2       \cr
{\displaystyle \frac{6}{5}}\:a_{31}\delta  \epsilon^5 &
       \sqrt{2}\:a_{32}\delta  \epsilon^2 &
                                                        a_{33}    \cr}
                                  \right)
\nonumber \\
\noalign{\bigskip}
Y_{\nu}(M_{GUT}) & = & \left(\matrix{
     \sqrt{2}\:a_{11}^{\prime\prime}\delta^3\epsilon^5 & 
                     2 \:a_{12}\delta  \epsilon^3 & 
                                                       0     \cr
       \mbox{\rule[-8mm]{0mm}{17mm}}
                                                       0     & 
{\displaystyle \sqrt{\frac{193}{670}}}\:a_{22}^{\prime}\delta^2\epsilon^2 &
                                      2 \:a_{23}\delta             \cr
                                                       0     & 
 {\displaystyle \frac{6}{5}}\:a_{32}^{\prime}\delta^2\epsilon^2 &
                                                       r_{\nu}a_{33}     \cr}
                              \right)
\nonumber \\
\noalign{\bigskip}
\frac{M_{RR}(M_{GUT})}{M_{R}} & = & \left(\matrix{
                     \phantom{\sqrt{2}}   A_{11}\delta  \epsilon^8 & 
                     \phantom{\sqrt{2}}   A_{12}\delta  \epsilon^6 & 
                     \phantom{\sqrt{2}}   A_{13}\delta  \epsilon^4 \cr
       \mbox{\rule[-5mm]{0mm}{12mm}}
                     \phantom{\sqrt{2}}   A_{12}\delta  \epsilon^6 & 
                     \phantom{\sqrt{2}}   A_{22}\delta  \epsilon^4 & 
                     \phantom{\sqrt{2}}   A_{23}\delta  \epsilon^2 \cr
                     \phantom{\sqrt{2}}   A_{13}\delta  \epsilon^4 & 
                     \phantom{\sqrt{2}}   A_{23}\delta  \epsilon^2 & 
                     \phantom{\sqrt{2}}                A_{33}     \cr}
                                      \right)
\nonumber
\eea
\caption{\label{GUT} The quark and lepton Yukawa matrices and neutrino Majorana
mass matrix as used in the analysis. In our numerical analysis
we set $ M_{R}=3\cdot 10^{14}$~GeV.
The Yukawa matrices follow from Eq.\ref{SRNDmRLopMtr} and the 
Clebsch factors arising from each operator are shown numerically
above. Clebsch zeroes play an important part in suppressing
the leading operator contribution in a particular element of the
matrix, or in simply giving a zero if all the operators are suppressed.
The Clebsch coefficients in the Majorana sector are set equal to unity, with 
$A_{ij}$ being independent order unity coefficients.}
\end{table}

In the neutrino sector the matrices in Table~\ref{GUT} satisfy the
condition of sequential dominance \cite{SRHND} in which a
neutrino mass hierarchy naturally results with the
dominant third right-handed neutrino being mainly responsible
for the atmospheric neutrino mass, and the sub-dominant second
right-handed neutrino being mainly responsible for the solar neutrino mass.
The atmospheric mixing angle is then determined approximately
as a ratio of $Y_{\nu}^{23}:Y_{\nu}^{33}$, and the solar
mixing angle is determined by a ratio of $Y_{\nu}^{12}$
to a linear combination of $Y_{\nu}^{22}$ and $Y_{\nu}^{32}$,
while the CHOOZ angle is determined by a more complicated formula
\cite{King:2002nf}. Note that the dominant right-handed neutrino
in this model is the heaviest one, corresponding to heavy
sequential dominance (HSD) and LFV has been considered
in general in this class of models \cite{Blazek:2002wq}. 

In the previous analysis \cite{BlKi_1} the matrix elements, 
$Y_{e}^{12}$, $Y_{e,\,\nu}^{13}$ were suppressed artificially to keep 
$BR(\mu\rightarrow e \gamma)$ within its experimental limit without substantially changing 
the predictions of fermion masses and mixings.  In this new analysis we have built 
this suppression into the model with our new choice of operators, 
whose Clebsch 
coefficients give zeros in the desired matrix elements as can be seen
in Table~\ref{GUT}. 
This can be understood analytically from \cite{Blazek:2002wq}. 

The subleading operators in the 33 position are
not shown explicitly, but are expected to lead to significant
deviations from exact Yukawa unification. This effect is parametrised
by the ratios
\beq
r_t\equiv \frac{Y_u(M_{GUT})_{33}}{Y_e(M_{GUT})_{33}},\ \ 
r_b\equiv \frac{Y_d(M_{GUT})_{33}}{Y_e(M_{GUT})_{33}}, \ \ 
r_{\nu}\equiv \frac{Y_{\nu}(M_{GUT})_{33}}{Y_e(M_{GUT})_{33}}.
\eeq

\section{Numerical Procedure\label{procedure}}

In our numerical analysis we have adopted a complete top-down approach
\cite{BCRW}. 
At the GUT scale the MSSM gauge couplings are related to the GUT scale 
couplings as 
$\alpha_{2L}=\alpha_{1}=\alpha_{GUT}$ and 
$\alpha_{3}=\alpha_{GUT}(1+\epsilon_3)$, 
where $\epsilon_{3}$ sums up the effects of GUT scale threshold corrections.  
The particular choice of the Yukawa couplings, Table \ref{GUT}, follows 
from the higher dimensional operators in Eq.~(\ref{SRNDmRLopMtr}) as the
latter are matched to the MSSM lagrangian. 
The parameters 
\beq
 \begin{array}{l}
     \mbox{\rule[-5mm]{0mm}{6mm}}
     M_{GUT},\;\alpha_{GUT},\;\epsilon_3,\;\delta,\;\epsilon,\; 
          a\mbox{'s}\;\,\mbox{and}\;\,A\mbox{'s},\; r_t,\; r_b,\; r_\nu,\\
     M_{1/2},\; A_0,\;\mu,\; B\mu,\; m_F^2,\; m_{\olbar{F}}^2,
                                  \; m_h^2\;\mbox{and}\;D^2
 \end{array}
 \label{GUT.input.1}
\eeq
are then defined by the boundary conditions at the 
GUT scale. They parametrise the imprint of a complete Pati-Salam theory
together with the SUSY sector (second line) on the MSSM and stand for 
the input in case this theory is not fully known. 
In the SUSY sector, the soft SUSY breaking parameters are 
for simplicity introduced at the same scale. The gaugino masses 
are assumed universal (equal to $M_{1/2}$) and so do 
the trilinear couplings: $A_i = A_0\,Y_i$, for $i=u,\:d,\:e,\:\nu$.
The soft scalar masses of the MSSM superfields include the $D$ terms from the
breaking of the Pati-Salam gauge group \cite{KiOl_yu}
\bea
   m_Q^2     &=& m_F^2 +  g_4^2\,D^2                        \nonumber\\
   m_{u_R}^2 &=& m_{\olbar{F}}^2 - (g_4^2-2g_{2R}^2) \, D^2 \nonumber\\
   m_{d_R}^2 &=& m_{\olbar{F}}^2 - (g_4^2+2g_{2R}^2) \, D^2 \nonumber\\
   m_L^2     &=& m_F^2 - 3g_4^2\,D^2 \\
   m_{e_R}^2 &=& m_{\olbar{F}}^2 +(3g_4^2-2g_{2R}^2) \, D^2 \nonumber\\
 m_{\nu_R}^2 &=& m_{\olbar{F}}^2 +(3g_4^2+2g_{2R}^2) \, D^2 \nonumber\\
   m_{H_u}^2 &=& m_{h  }^2 -        2g_{2R}^2  \, D^2       \nonumber\\
   m_{H_d}^2 &=& m_{h  }^2 +        2g_{2R}^2  \, D^2.      \nonumber
\label{eq:D}
\eea
As $D^2=\frac{1}{8}\left(\,|\olbar{H}_{\nu}|^2-|H_{\nu}|^2\,\right)$ 
\cite{KiOl_yu} it is possible for this quantity to be both 
positive and negative.

We now describe minor simplifications to the input in (\ref{GUT.input.1})
which were assumed in the actual numerical analysis. We have kept
equality between the two order parameters $\delta$ and $\epsilon$
as in Eq.(\ref{eq:de}) and the
soft SUSY breaking scalar masses $m_F$ and $m_{\olbar{F}}$ have been 
held equal to each other as well.
Furthermore we exploited the fact that determining $\mu(M_{GUT})$ and 
$B\mu$ at the GUT scale is equivalent to determining the
low energy values $\mu(M_Z)$ and  $\tan\beta$, respectively. 
Thus instead of (\ref{GUT.input.1}) our numerical analysis uses
\beq
 \begin{array}{l}
     \mbox{\rule[-5mm]{0mm}{6mm}}
     M_{GUT},\;\alpha_{GUT},\;\epsilon_3,\;\delta,\; 
          a\mbox{'s}\;\,\mbox{and}\;\,A\mbox{'s},\; r_t,\; r_b,\; r_\nu,\\
     M_{1/2},\; A_0,\;\mu(M_Z),\; \tan\beta,\; m_F^2,
                                  \; m_h^2\;\mbox{and}\;D^2
 \end{array}
 \label{GUT.input.2}
\eeq
as input parameters.
The top down approach implies that we can freely
vary or hold fixed any one of them and then investigate the fit properties. 
This is one of the advantages of doing the analysis top down. 
For example, in more traditional bottom up approaches it is 
difficult to control the size of the dimensionless GUT scale parameters.
One usually sets up a sample of randomly scattered points and then 
searches through it to identify a sub-sample with physically interesting 
GUT scale properties.
In our case we can set up the interesting GUT relations explicitly right
at the start --- as we have done for instance in section \ref{yuk} where
the fits with $r_b$ and $r_t$ approaching unity are studied.

We note that taking advantage of the top-down approach 
we kept $\delta=0.22$, $r_\nu=1$, $A_0=0$ and $\tan\beta=50$ fixed 
throughout the analysis. This effectively increases the degrees of freedom
of the global analysis and can provide a reference point if further analysis
is required in the future. 
We also kept the $\mu$ parameter at scale $M_Z$ fixed to two different 
values as is explained below.

Two-loop RGEs for the dimensionless couplings and 
one-loop RGEs for the dimensionful couplings were used to 
run all couplings down to the scale $M_{3R}$ where the heaviest
right-handed neutrino decoupled from the RGEs. Similar steps
were taken for the lighter $M_{2R}$ and $M_{1R}$ scales, 
and finally with all three right-handed neutrinos decoupled
the solutions for the MSSM couplings and spectra were computed 
at the $Z$ scale. This includes full one loop SUSY threshold corrections 
to the fermion mass matrices and all Higgs masses
while the sparticle masses are obtained at tree level.

$m_h$ and $D$ in Eqs.~(\ref{eq:D}) 
were varied to optimise radiative electroweak symmetry breaking (REWSB), 
which was checked at one loop with the leading $m_t^4$ and $m_b^4$
corrections included
following the effective potential method in \cite{EPM}.
We note that as $\tan\beta$ determines the Higgs bilinear parameter $B\mu$,
there is a redundancy in our procedure since two input parameters, 
$m_h$ and $D$, determine one condition for the Higgs VEV of $246\,$GeV.
This approach enabled us to control the $\mu$ parameter
and we explored regions with $\mu$ low 
($\mu = 120$GeV) and high ($\mu = 300$GeV) \footnote
{
For $\tan\beta$ as large as 50,   $\mu\gg 300$GeV 
leads to too large SUSY threshold corrections to the
masses of the third generation fermions $\tau$ and $b$
unless the sparticles in the loop have masses well above 
the $1$ TeV region.
\cite{large.mb,BCRW}
}.

An experimental lower bound on each sparticle mass was imposed. 
In particular,
the most constraining are: the LEP limits on the charged SUSY masses
($m_{\tilde{\chi}^\pm},m_{\tilde{\tau}}>105$GeV), the CDF limit
on the mass of the $CP$ odd Higgs state 
($m_{A^0}>105$-$110\,$GeV, valid for $\tan\beta\approx 50$) \cite{Tevatron}, 
and the requirement that the lightest SUSY particle should be neutral.  
Finally, the $\chi^2$ function $\;\sum (X_i^{th}-X_i^{exp})^2/\sigma_i^2)\;$
is evaluated based on the agreement between the theoretical predictions
and 24 experimental observables collected in Table \ref{t:observables}.
In addition to the constraints listed above and in \cite{BlKi_1}, 
we make a full analysis of the quark sector mass and mixings, in 
particular we have included the important constraint set by 
BR($b \rightarrow s \gamma$).

%%%%%\newpage

\begin{table}
\begin{center}
\begin{tabular}{ccc}
%GG \= bbbbbbbbbbbbb \= eeeeeeeeeeeeeeee \= aaaaaaaaaaaaaa \kill
Observable & { Mean      } & {$\sigma_{ i}$} \rule[-0.60cm]{0mm}{0.7cm}\\
\noalign{\hrule height\rulerheight}
\noalign{\medskip}
      $\alpha_{EM}  $    & $1/137.036$            & ${7.30\cdot 10^{-6}}$ \\ 
      $G_\mu     $    & $1.16639\cdot 10^{-5}$ & ${1.12\cdot 10^{-7}}$ \\
      $\alpha_s(M_Z)$    & $0.1181$               & $0.0020           $ 
\rule[-0.60cm]{0mm}{0.7cm}    \\
        $M_t$  & $174.3$ & $5.1$ \\
        $m_b(m_b)  $    & $4.20  $               & ${0.20}        $ \\
       $M_b-M_c   $    & $3.4   $               & ${0.2}         $ \\
   $m_s(2\mbox{GeV})$   & $0.110 $               & ${0.035}       $ 
\rule[-0.40cm]{0mm}{0.7cm}                                                 \\
$(m_d^2-m_u^2)/m_s^2$ & $2.03\cdot 10^{-3}$    & ${2.0\cdot 10^{-4}}$ 
\rule[-0.40cm]{0mm}{0.7cm}                                                 \\
       $m_d/m_s$       & $0.05  $               & ${0.015}        $ \\
        $M_\tau $       & $1.777 $               & ${1.8\cdot 10^{-3}} $ \\
% zzz !!! zzz
        $M_\mu  $       & $0.106 $               & ${1.1\cdot 10^{-4}} $ \\
        $M_e    $       & $5.11\cdot 10^{-4}$    & ${5.1\cdot 10^{-7}} $ \\
        $|V_{us}| $     & $0.2196 $              & $0.0023           $ \\
        $|V_{cb}| $     & $0.0402 $              & $0.003           $ \\
  $|V_{ub}|/|V_{cb}|$   & $0.09   $              & $0.02             $ \\
%        $\eps_K   $     & $2.28\cdot 10^{-3}$   & ${0.23\cdot 10^{-3}}$ 
%  
\rule[-0.60cm]{0mm}{0.7cm}                                                 \\
        $M_Z      $     & $91.1882$    & ${0.091}        $ \\
        $M_W      $     & $80.419  $   & ${0.08}        $ \\
       $\rho_{NEW}$     & $-0.0002 $   & $0.0011           $  \\
   $BR(b\rightarrow s\gamma)$& $3.47\cdot 10^{-4}$ & $0.45\cdot 10^{-4}$ \\
       $\delta a_{\mu\;NEW}$     & $34.7.6\cdot 10^{-10}$   & 
                                      $11\cdot 10^{-10}$ 
\rule[-0.60cm]{0mm}{0.7cm}                                                 \\
       $\Delta m^2_{ATM}$   & $2.5\cdot 10^{-3}$    & 
                                       $ 0.8\cdot 10^{-3}        $ \\
       $\sin^22\theta_{ATM}$& $0.99$                 & $0.06$ \\
       $\Delta m^2_{SOL}$   & $7.0\cdot 10^{-5}$    & 
                                       $ 3\cdot 10^{-5}        $ \\
       $\sin^22\theta_{SOL}$& $0.8$                 & $0.09$ \\
\vspace{0.3cm}\\
\noalign{\hrule height\rulerheight}
\noalign{\medskip}

\end{tabular}
\vskip10mm
\begin{minipage}[t]{15cm}
\caption{
{\small Table of observables and $\sigma$ values which are used to calculate the $\chi^2$ that enables best fit regions to be determined via minimisation.}}\label{t:observables}
\end{minipage}
\end{center}
\end{table}
\newpage

\section{Results and Discussion}\label{res_dis}

The numerical results from the global analysis are presented 
in the form of contour plots in the $(m_F,M_{1/2})$ plane and 
are produced for two different values of the mu parameter 
$\mu = 120\,$GeV and 
$\mu = 300\,$GeV.
Before we address the details we would like to discuss two different 
viewpoints of our analysis, namely the flavour sector on the one
hand and the unification sector of the other hand.
In our discussion we would like to distinguish between the two
viewpoints. The main distinction is that in the MSSM analysis the
flavour parameters 
$a_{ij}$ (with the exception of $a_{33}$) and $A_{ij}$ can be considered 
fixed at unity or at a value of order unity. Up to $a_{33}$ which enters
the large Yukawa couplings their exact values do not 
affect the fit of the SUSY spectra or SUSY-related observables like the muon 
$g-2$ or branching ratio $b\to s \gamma$. They neither perturb gauge 
coupling unification nor change the running of the large Yukawa couplings.
This means that the discussion of our results is naturally split into a part
dealing with the flavour structure of the Pati-Salam model 
where the variation of the coefficients of the higher dimensional 
operators matters, and a part where the 
MSSM analysis is presented and the conclusions do not depend on
the variation of the $a$ and $A$ parameters (up to $a_{33}$).

Concerning the flavour sector,
our results can be used to show how well the model, 
i.e. the set of higher dimensional operators specified by 
Eq.~(\ref{SRNDmRLopMtr}), describes the observed fermion masses and mixings.
Taking this viewpoint all parameters listed in (\ref{GUT.input.2}) 
represent the input of the analysis. The results in either of the
four panels in Figure~\ref{1} show that the model gives a very good
agreement with the data. The minimum of the total $\chi^2$ is less
than unity obtained for $\mu=120\,$GeV in the upper left panel.
This means that it is possible to fit every observable to better than
a $1\,\sigma$ accuracy. 

Concerning the unification sector, the conclusions are much
stronger as much fewer number of the input parameters enters
effectively after the $a$'s and $A$'s decouple from the analysis.
Indeed, the set of the effective input parameters in this sector 
is reduced to
\beq
 \begin{array}{l}
     \mbox{\rule[-5mm]{0mm}{6mm}}
     M_{GUT},\;\alpha_{GUT},\;\epsilon_3,\; 
          a_{33},\; r_t,\; r_b,\\
%     M_{1/2},\; \mu(M_Z),\;  m_F^2,
     M_{1/2},\;  m_F^2, \; m_h^2\;\mbox{and}\;D^2.
 \end{array}
 \label{GUT.input.3}
\eeq
With this input the low energy Higgs and SUSY spectra are determined.
The conventional present-day observables include 
$\alpha_{EM}$, $G_\mu$, $\alpha_s(M_Z)$, $M_t$, $m_b(m_b)$,
$M_\tau$, $M_Z$, $M_W$, $\rho_{NEW}$, $BR(b\to s \gamma)$ and 
$\delta a_\mu$. 
The $a_{ij}$ and $A_{ij}$ input parameters are all of order one
and their exact values are always adjusted to fit the first two 
generation masses and mixings well while these variations do not 
affect the fit of the observables listed above. 

We study many details of the MSSM analysis, in particular the dependence
of the fit on $m_F^2$ and $M_{1/2}$, best fit results for muon $g-2$,
and $BR(\tau \to \mu\gamma)$ predictions.
The numerical results also contain studies of a deviation from Yukawa 
unification and a future measurement of $BR(\tau \rightarrow \mu \gamma)$.  
The effect of a change to the present muon $g-2$ discrepancy was studied 
and also the effect of future direct Higgs searches, the results of which 
can also be found at the end of the paper.

\begin{table}
\begin{center}
\begin{tabular}{|l||r|r|r|r|}
\noalign{\hrule height\rulerheight}
\multicolumn{5}{|c|}{\bf Inputs}\\
\noalign{\hrule height\rulerheight}
& \multicolumn{2}{|c|}{\bf $\mu=120$~GeV}& \multicolumn{2}{|c|}{\bf $\mu=300$~GeV}\\
\noalign{\hrule height\rulerheight}
        &   Min A   &   Min B&   Min A   &   Min B \\
\noalign{\hrule height\rulerheight}
$M_{1/2}            $&$450             $&$  650             $&$450             $&$  650          $\\
$m_{F}              $&$500             $&$  650             $&$500             $&$  650          $\\
$\mu                $&$ 120            $&$ 120              $&$ 300            $&$ 300           $ \\
$D^2                $&$-6.4\cdot 10^{4}$&$  17\cdot 10^{4}  $&$ -10\cdot 10^{4}$&$  13\cdot10^{4}$\\   
$m_{h}^{2}          $&$  6\cdot 10^{5} $&$  16\cdot 10^{5}  $&$ 4.5\cdot 10^{5}$&$ 14\cdot 10^{5}$\\ 
\noalign{\hrule height\rulerheight}
$ r_{t}             $&$  1.01          $&$  1.07            $&$ 1.03           $&$  1.02         $\\ 
$ r_{b}             $&$  0.75          $&$0.72              $&$ 0.66           $&$  0.64         $\\
$  a_{33}           $&$  0.55          $&$ 0.55             $&$ 0.55           $&$   0.56        $\\ 
\noalign{\hrule height\rulerheight}
$  a_{11}           $&$  -0.93       $&$  -0.92     $&$-0.92       $&$-0.93        $\\ 
$  a_{12}           $&$  0.20        $&$ 0.33       $&$ 0.31       $&$ 0.30        $\\ 
$  a_{21}           $&$  1.67        $&$ 1.67       $&$ 1.67       $&$ 1.75        $\\ 
$  a_{22}           $&$  1.13        $&$ 1.12       $&$ 1.13       $&$ 1.13        $\\ 
$  a_{23}           $&$  0.98        $&$ 0.89       $&$ 1.05       $&$ 0.85        $\\ 
$  a_{31}           $&$  -0.20       $&$ -0.21      $&$-0.20       $&$-0.28        $\\ 
$  a_{32}           $&$  2.18        $&$ 2.08       $&$ 2.32       $&$ 2.53        $\\ 
\noalign{\hrule height\rulerheight}
$  a_{12}^{\prime}    $&$  0.77         $&$0.77        $&$ 0.71          $&$0.71           $\\ 
$  a_{13}^{\prime}    $&$  0.60         $&$ 0.53       $&$ 0.46          $&$0.46           $\\ 
$  a_{22}^{\prime}    $&$  0.66         $&$  0.66      $&$ 0.64          $&$0.62           $\\ 
$  a_{23}^{\prime}    $&$  0.41         $&$ 0.40       $&$ 0.36          $&$0.36           $\\ 
$  a_{32}^{\prime}    $&$  1.16         $&$ 1.80       $&$ 1.56          $&$1.72           $\\ 
\noalign{\hrule height\rulerheight}
$a_{11}^{\prime\prime}$&$  0.32         $&$ 0.278     $&$   0.20        $&$   0.23        $\\ 
\noalign{\hrule height\rulerheight}
$A_{11}$&$  0.63        $&$ 0.94   $&$ 0.63        $&$  0.94     $\\ 
$A_{12}$&$  0.74        $&$ 0.48   $&$ 0.69        $&$  0.52     $\\ 
$A_{13}$&$  1.75        $&$ 2.10   $&$ 1.73        $&$  2.04     $\\ 
$A_{22}$&$ 0.97         $&$ 0.52   $&$ 0.93        $&$  0.55     $\\ 
$A_{23}$&$ 2.49         $&$ 1.79   $&$ 2.23        $&$  1.91     $\\ 
$A_{33}$&$ 1.97         $&$ 1.88   $&$ 1.97        $&$  1.88     $\\ 
\noalign{\hrule height\rulerheight}
\end{tabular}
\begin{minipage}[t]{15cm}
\caption{\small Tables of inputs for the best fit points for each of the global $\chi^2$ minima with
$\mu=120$ and $300$~GeV.}\label{input}
\end{minipage}
\end{center}
\end{table}

\begin{table}
\begin{center}
\begin{tabular}{|l||r|r|r|r|}
\noalign{\hrule height\rulerheight}
\multicolumn{5}{|c|}{\bf Outputs}\\
\noalign{\hrule height\rulerheight}
& \multicolumn{2}{|c|}{\bf $\mu=120$~GeV}& \multicolumn{2}{|c|}{\bf $\mu=300$~GeV}\\
\noalign{\hrule height\rulerheight}
        &   Min A   &   Min B&   Min A   &   Min B \\
\noalign{\hrule height\rulerheight}
$m_{A^0}       $&$102     $&$  818    $&$102     $&$ 822   $   \\
$m_{h^0}       $&$106     $&$  114    $&$106     $&$ 114   $   \\
$m_{H^0}       $&$112     $&$  891    $&$113     $&$ 888   $  \\
$m_{H^+}       $&$136     $&$  861    $&$135     $&$ 861   $   \\
\noalign{\hrule height\rulerheight}
$M_{1}              $&$186$&$   270   $&$ 186     $&$ 271  $\\
$M_{2}              $&$371$&$   537   $&$ 371     $&$ 537  $\\
$M_{3}      $&$1175$&$  1671  $&$ 1175    $&$ 1671 $ \\
\noalign{\hrule height\rulerheight}
$M_{\chi^{+}_{1}}   $&$114 $&$   117 $&$272$&$ 290   $ \\
$M_{\chi^{+}_{2}}   $&$ 390$&$   549 $&$408$&$ 554   $ \\
$M_{\tilde{N}_{1}}  $&$ 98 $&$   107 $&$179$&$  249  $\\
$M_{\tilde{N}_{2}}  $&$130 $&$   127 $&$277$&$  305  $\\
$M_{\tilde{N}_{3}}  $&$ 198$&$   278 $&$307$&$  311  $ \\
$M_{\tilde{N}_{4}}  $&$390 $&$   549 $&$408$&$  554  $\\
\noalign{\hrule height\rulerheight}
$M_{\tilde{Q}_{ 1,\,2}}$&$ 1166   $&$ 1679  $&$ 1159   $&$ 1673  $ \\       
$M_{\tilde{Q}_{ 3}}    $&$ 979    $&$  1345 $&$ 960    $&$ 1356  $\\
$M_{\tilde{U}_{ 1,\,2}}$&$ 1131   $&$ 1623  $&$ 1124   $&$ 1617  $ \\      
$M_{\tilde{U}_{ 3}}    $&$ 798    $&$  1147 $&$ 805    $&$ 1160  $\\
$M_{\tilde{D}_{ 1,\,2}}$&$ 1182   $&$ 1510  $&$ 1204   $&$ 1529  $ \\
$M_{\tilde{D}_{ 3}}    $&$ 923    $&$ 1192  $&$ 1044   $&$ 1251  $\\
\noalign{\hrule height\rulerheight}
$M_{\tilde{L}_{ 1}}    $&$ 673 $&$ 611      $&$715  $&$  656    $\\ 
$M_{\tilde{L}_{ 2}}    $&$ 665 $&$ 595      $&$707  $&$  644    $\\  
$M_{\tilde{L}_{ 3}}    $&$ 580 $&$ 334      $&$638  $&$  425    $\\ 
$M_{\tilde{E}_{ 1}}    $&$ 496 $&$ 766      $&$473  $&$  752    $\\
$M_{\tilde{E}_{ 2}}    $&$ 495 $&$ 765      $&$473  $&$  751    $\\
$M_{\tilde{E}_{ 3}}    $&$ 201 $&$ 370      $&$188  $&$  325    $\\
\noalign{\hrule height\rulerheight}
$\tau\rightarrow\mu\gamma$&$ 2\cdot 10^{-7} $  &$   3\cdot 10^{-6}$&$ 8\cdot 10^{-8} $&$ 5\cdot 10^{-7}   $\\
$\tau\rightarrow e\gamma $&$ 1 \cdot 10^{-14} $&$  3\cdot 10^{-13}$&$ 6\cdot 10^{-15} $&$ 5\cdot 10^{-14} $ \\
$\mu\rightarrow e\gamma  $&$ 3\cdot 10^{-14} $ &$ 1\cdot 10^{-13} $&$ 1\cdot 10^{-14} $&$ 3\cdot 10^{-14} $\\
\noalign{\hrule height\rulerheight}
$\sin\theta_{13}         $&$ 0.053   $&$ 0.078             $&$  0.037   $&$     0.10              $\\
\noalign{\hrule height\rulerheight}
$\sin(\beta-\alpha)$&$0.22     $&$1.0     $&$ 0.15    $&$  1.0  $   \\
$\cos(\beta-\alpha)$&$-0.98    $&$0.0     $&$ -0.99    $&$ 0.0   $   \\
\noalign{\hrule height\rulerheight}
\end{tabular}
\begin{minipage}[t]{15cm}
\caption{\small Tables of outputs for the best fit points for each of the global $\chi^2$ minima with
$\mu=120$ and $300$~GeV. The input parameters are as defined in Table~\ref{input}.}\label{output}
\end{minipage}
\end{center}
\end{table}

From our global analysis we found that there are two $\chi^2$ minima
as shown in Figure~\ref{1}.
In this model there are two conditions and three free variables, 
$m_{h}^2,\,D^2,$ and $B\mu$, for electroweak symmetry breaking to be 
achieved. The two minima hence are independent solutions to these 
conditions. Minimum A has $D^2$ negative and smaller $m_{h},\,B\mu$. 
Minimum B on the other hand has $D^2$ positive and larger $m_{h},\,B\mu$. 
The relative size of $B\mu$ results in a different Higgs spectrum, 
particularly the CP odd pseudoscalar Higgs, $A^{0}$, which will be 
lighter for minimum A and heavier for minimum B. The difference between 
the sign of $D^2$, which contributes to the soft scalar masses as shown 
in Eq.~(26), means that minimum B will have lighter right squarks and left 
sleptons, along with heavier left squarks and right sleptons, than minimum A.
This difference in sign of $D^2$ has some interesting phenomenological 
consequences for the two minima which will now be discussed.

The upper and lower plots shown in Figure~\ref{1}, display the $\chi^2$ 
contours for these two minima.  Each of the figures display results for 
both $\mu=120$ and $300$~GeV in the relative left and right positions.  
The contours in Figure~\ref{1} are bounded from the lower $m_F$ region due 
to the lightest stau becoming the LSP and from the lower $M_{1/2}$ region 
due to an increasing $\chi^2$ penalty coming from 
$BR(b \rightarrow s \gamma)$.  

The upper minima of Figure~\ref{1}, Minima A, have a preferred region in the 
lower $(m_F,M_{1/2})$ plane, with $M_{1/2}=400\,-\, 500$~GeV and 
$m_F=500\,-\, 700$~GeV.  The lower minima of Figure~\ref{1}, Minima B, have 
their preferred region nearer $M_{1/2}=550\,-\, 650$~GeV and 
$m_F=600\,-\, 800$~GeV. A list of inputs and outputs for the best fit
point in each minimum can be found in Tables~\ref{input} and 
\ref{output}. The Higgs masses and CP even Higgs mixings 
found for minimum A in Table~\ref{output} are
discussed in detail in section~\ref{Higgsmass}.

\subsection{Muon $g-2$}

\begin{figure}[p]
\begin{center}
\scalebox{0.8}{\includegraphics*{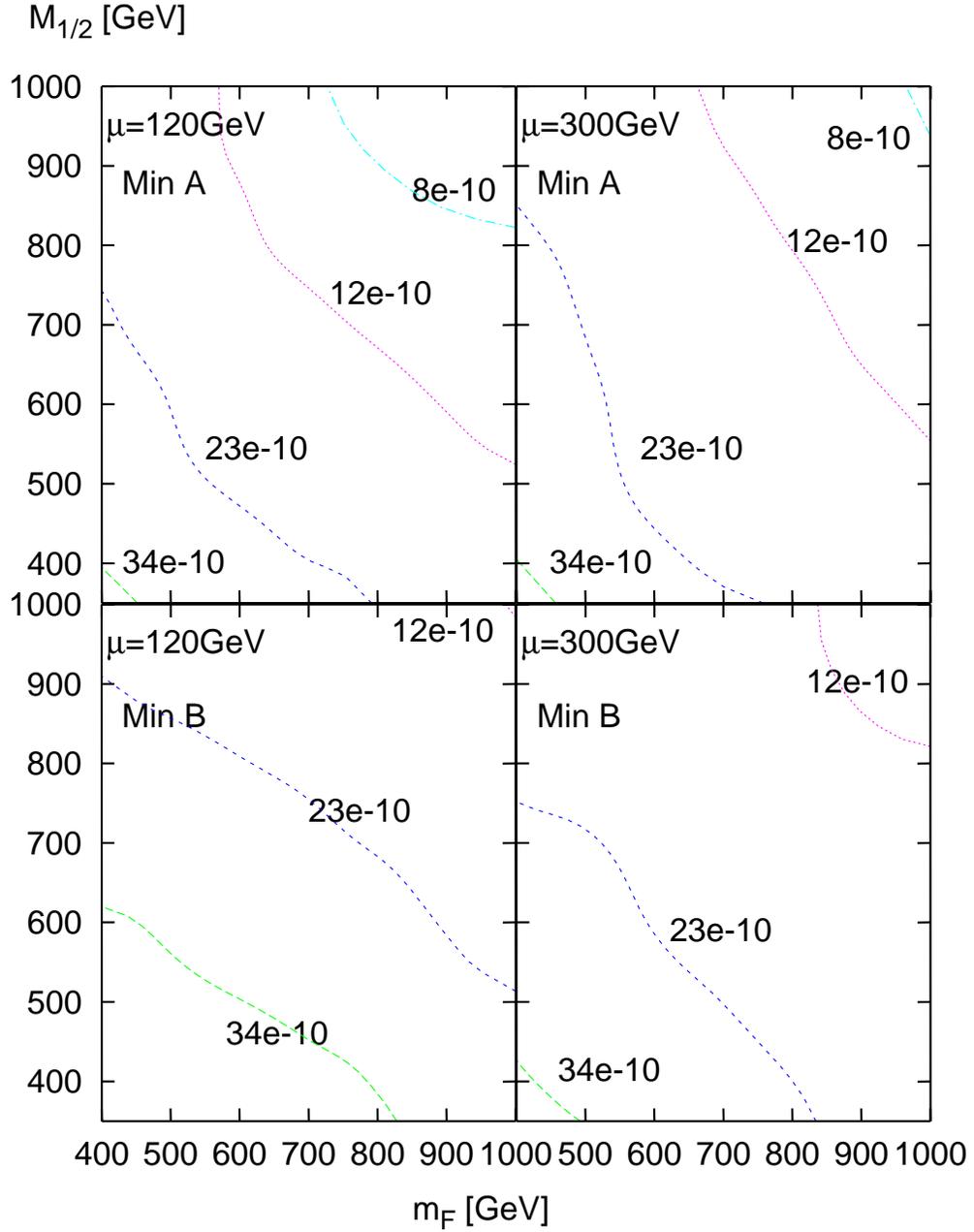}}
\vskip-20mm
\begin{minipage}[t]{15cm}
\caption{\small{Muon $g-2$ contour plot in the plane of $(m_F,M_{1/2})$.  The four plots, are obtained from the two minima, minimum A and minimum B with $\mu=120$ and $300$~GeV as labelled.  All points to the left of the solid red line are unphysical due to the lightest stau becoming the LSP. The present discrepancy stands at $34(11)\times 10^{-10}$ with the above plots showing 1 and 2 $\sigma$contours.}}\label{2}
\end{minipage}

\end{center}
\end{figure}

\begin{figure}[p]
\begin{center}
\scalebox{0.8}{\includegraphics*{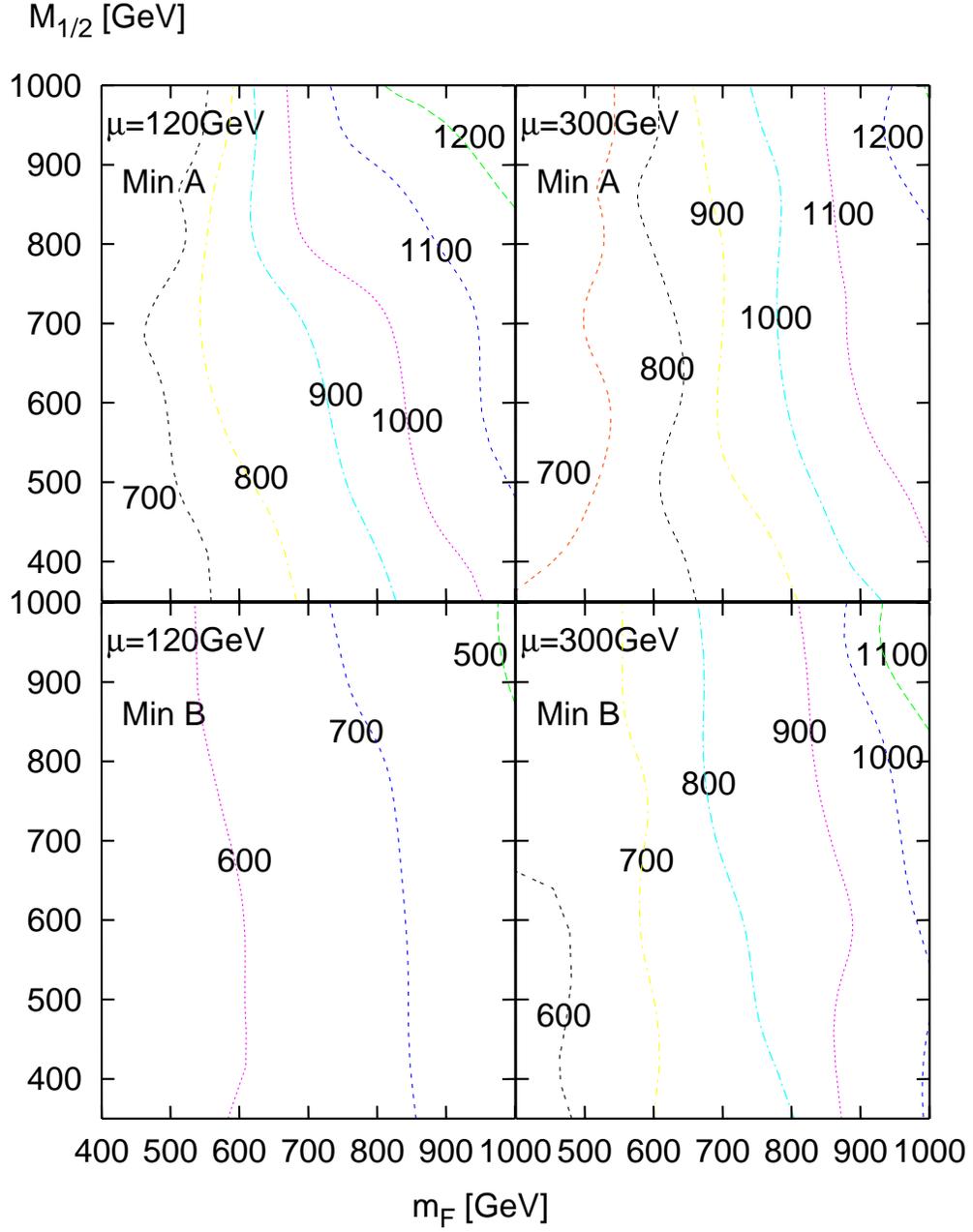}}
\vskip-20mm
\begin{minipage}[t]{15cm}
\caption{\small{Contours of the second generation sneutrino mass, $m_{\tilde{\nu}_{\mu}}$, are plotted in the plane of $(m_F,M_{1/2})$. The contours are in the units of~GeV.  The four plots, are obtained from the two minima, minimum A and minimum B with $\mu=120$ and $300$~GeV as labelled.  All points in the top left corner with approximately $M_{1/2}> 700$~GeV and $m_F> 700$~GeV are unphysical due to the lightest stau becoming the LSP.}}\label{13b}
\end{minipage}

\end{center}
\end{figure}

Figure~\ref{2}, shows contour plots for the SUSY contributions towards the muon $g-2$.  Both minimum A and minimum B (upper and lower plots respectively) give good fits to the present discrepancy between experiment and Standard Model prediction.  As expected, a larger contribution to the muon $g-2$ is obtained in the lower left corner of the $(m_F,M_{1/2})$ plane where the SUSY spectrum is lightest and decreases as we move towards a heavier spectrum in the top right corner. It is also clear that for any one point in the $(m_F,M_{1/2})$ plane, minimum B gives a larger contribution than the corresponding point in minimum A.  This relative enhancement can be ascribed to the dominant chargino-sneutrino diagram via the presence of a lighter muon sneutrino for the case of minimum B, as can be seen in Figure~{\ref{13b}}.  

The present muon $g-2$ discrepancy lies at $34\times 10^{-10}$ but over the past 12 months it has varied from a $1.5\sigma$ to $3\sigma$ effect.  Also the size of the present discrepancy depends on the experimental data used in the calculation of the Standard model prediction.  The value we have used throughout our analysis \cite{muon_SM} makes use of $e^{+}e^{-}$ data.  On the other hand it is possible to do the same calculation making use if $\tau$ decay data \cite{g-2_tau}, which gives a lower discrepancy of $9.4\times 10^{-10}$.  As a result we think it worth while looking into how our best fit regions would change if a lower discrepancy was assumed.  For simplicity we took 3 points in the $(m_F,M_{1/2})$ plane of minimum A with $\mu=120$~GeV and gradually changed the $g-2$ discrepancy from $34\times 10^{-10}$ down to $0$.  The results are presented in Figure~{\ref{32a}} as a plot of $\chi^2$ against the muon $g-2$ discrepancy, $a_{\mu}^{New}$.  

With the discrepancy held at $34\times 10^{-10}$ the best fit point is near $M_{1/2}=450$~GeV and $m_F=550$~GeV. Following the curve corresponding to this point in parameter space, we can see that as the muon $g-2$ discrepancy is lowered the $\chi^{2}$ gradually increased. Therefore the best fit point has moved in the positive $M_{1/2}$, $m_F$ direction. Looking at the two further curves in Figure~\ref{32a} we can see that if $a_{\mu}^{New}\sim 16\times 10^{-10}$ then the best fit point would move nearer $M_{1/2}=550$~GeV and $m_F=650$~GeV. One particular point of interest is $a_{\mu}^{New}=9.4\times 10^{-10}$, the value for the discrepancy as given by the Standard Model prediction from $\tau$ decay data.  If we make an approximation, based on the curves in Figure~{\ref{32a}}, we can say that the best fit point, for $a_{\mu}^{New}=9.4\times 10^{-10}$, would be in the region $M_{1/2}=550-700$~GeV, $\,m_F=650-700$~GeV.

\subsection{$\tau \rightarrow \mu \gamma$}

\begin{figure}[p]
\begin{center}
\scalebox{0.8}{\includegraphics*{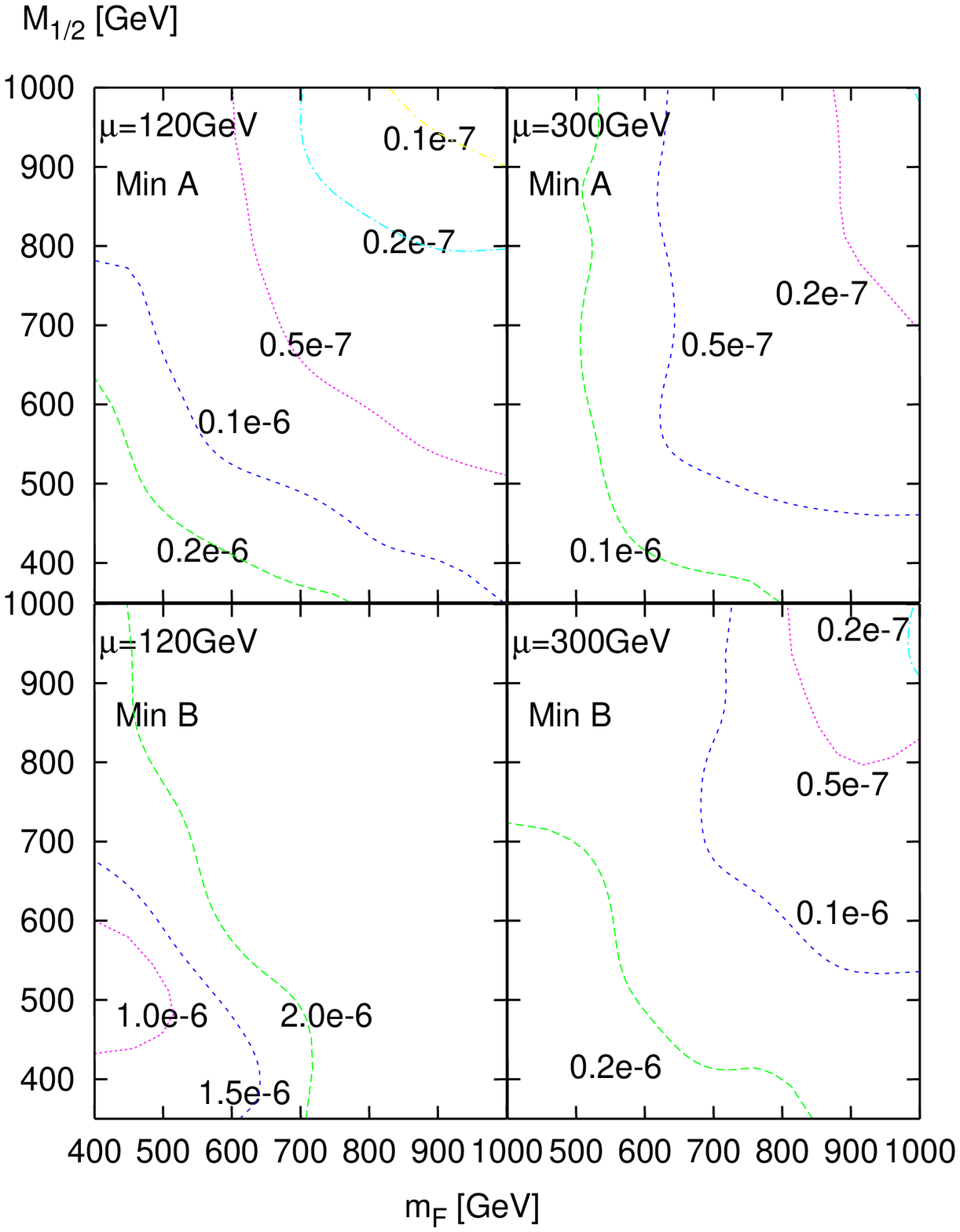}}
\vskip-20mm
\begin{minipage}[t]{15cm}
\caption{\small{Contours of $BR(\tau \rightarrow \mu \gamma)$ are plotted in the plane of $(m_F,M_{1/2})$.  The four plots, are obtained from the two minima, minimum A and minimum B with $\mu=120$ and $300$~GeV as labelled.  All points in the top left corner with approximately $M_{1/2}> 700$~GeV and $m_F> 700$~GeV are unphysical due to the lightest stau becoming the LSP.}}\label{5}
\end{minipage}

\end{center}
\end{figure}

Figure~\ref{5} displays contours for the quantity BR$(\tau\rightarrow\mu\gamma)$ for both minima with $\mu=120$ and $300$~GeV as labelled. The general pattern of the contours show larger branching ratio for lighter SUSY spectrum and smaller branching ratio for heavier spectrum. This pattern is not strictly obeyed in the bottom left panel which shows results for minimum B with $\mu=120$~GeV. The reason for this is that our numerical procedure adds a large penalty $\chi^2$ contribution for a $\tau\rightarrow \mu \gamma$ branching ratio larger than the BaBar limit of $2.0 \times 10^{-6}$. Looking at the bottom left panel in Figure~\ref{5} we would expect the branching ratio to exceed the present limit as we go to a lighter spectrum. The result of adding this penalty $\chi^2$ is to numerically force an alternative solution to be found which gives lower branching ratio and disrupts the pattern. Recalculation of this region of parameter space without the additional $\chi^2$ penalty does indeed yield values of BR$(\tau\rightarrow \mu \gamma)$ as large as $6.0 \times 10^{-6}$, these points would therefore follow the expected contour pattern but are clearly experimentally excluded.

Looking at Figure~\ref{5}, the branching ratio for minimum A with $\mu=120$ and $300$~GeV is well below the present experimental bound.  On the other hand, the branching ratio for minimum B, with $\mu=120$~GeV Figure~\ref{5}, is right at the present 90\% confidence level bound of $2.0 \times 10^{-6}$ \cite{babar_tmg}.  For $\mu=300$~GeV minimum B gives a branching ratio in the range, $0.1-0.2 \times 10^{-6}$, just below the present bound.  With BaBar expected to search as far as BR$(\tau\rightarrow\mu\gamma)\sim 10^{-8}$ over the next 5 years this certainly provides a means of distinguishing the two minima.

\subsection{Deviations from Yukawa Unification}\label{yuk}

\begin{figure}[p]
\begin{center}
\scalebox{0.8}{\includegraphics*{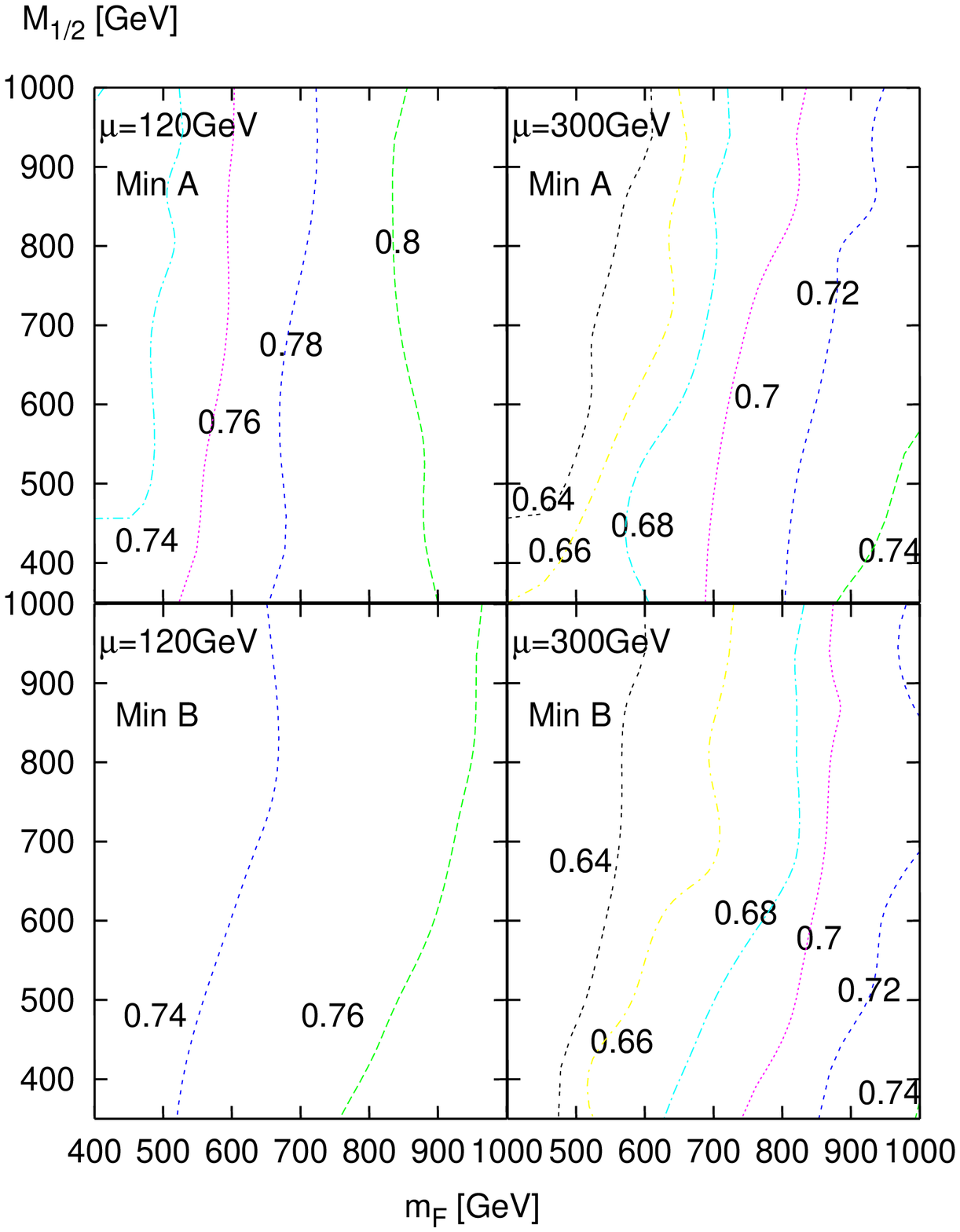}}
\vskip-20mm
\begin{minipage}[t]{15cm}
\caption{\small{Contours of $r_{b}=Y_{b}/Y_{\tau}$ are plotted in the plane of $(m_F,M_{1/2})$.  The four plots, are obtained from the two minima, minimum A and minimum B with $\mu=120$ and $300$~GeV as labelled.  All points in the top left corner with approximately $M_{1/2}> 700$~GeV and $m_F> 700$~GeV are unphysical due to the lightest stau becoming the LSP.}}\label{3}
\end{minipage}

\end{center}
\end{figure}

\begin{figure}[p]
\begin{center}
\scalebox{0.8}{\includegraphics*{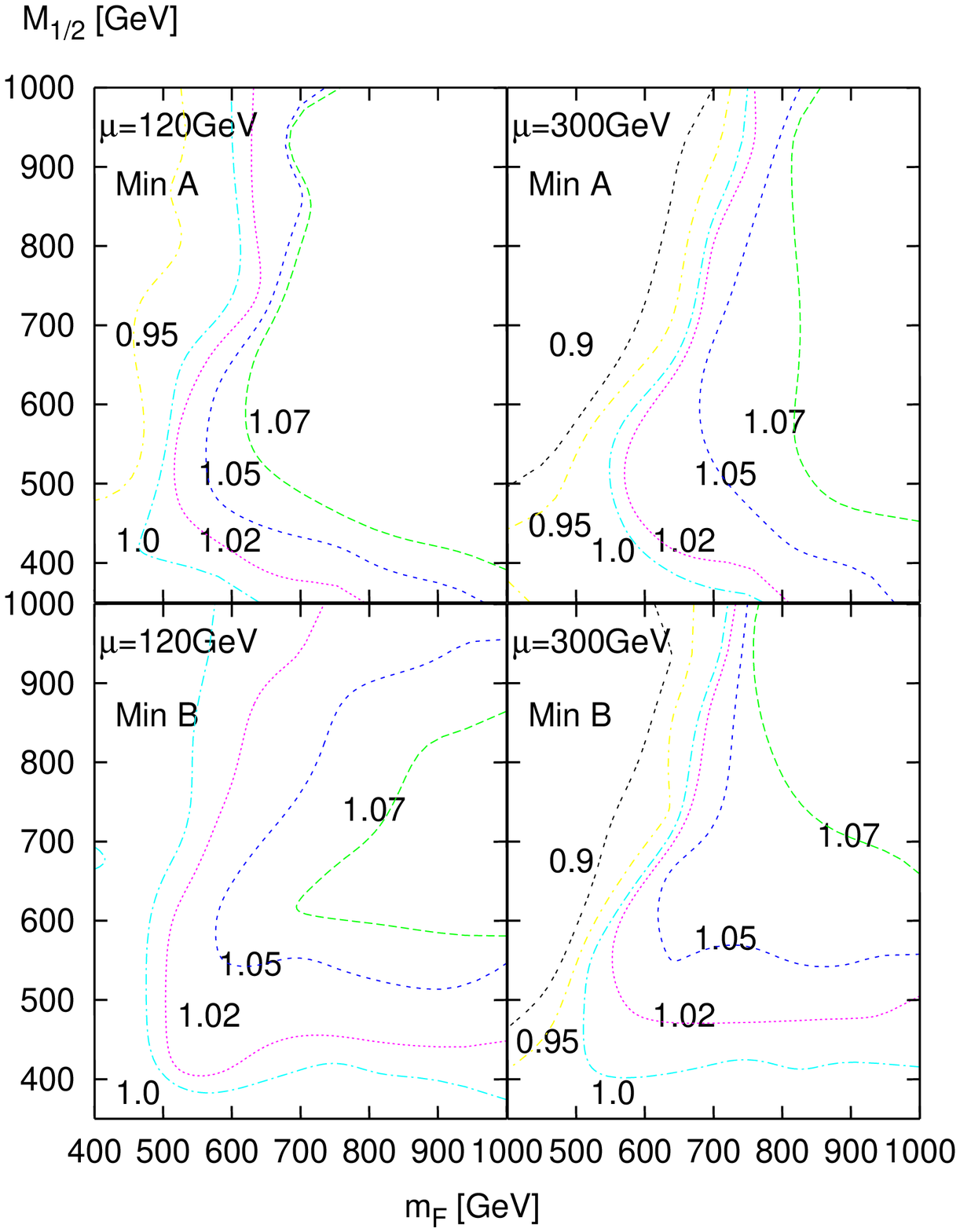}}
\vskip-20mm
\begin{minipage}[t]{15cm}
\caption{\small{Contours of $r_{t}=Y_{t}/Y_{\tau}$ are plotted in the plane of $(m_F,M_{1/2})$.  The four plots, are obtained from the two minima, minimum A and minimum B with $\mu=120$ and $300$~GeV as labelled.  All points in the top left corner with approximately $M_{1/2}> 700$~GeV and $m_F> 700$~GeV are unphysical due to the lightest stau becoming the LSP.}}\label{4}
\end{minipage}

\end{center}
\end{figure}

The plots shown in Figure~\ref{3} show contour lines for $r_{b}=Y_{b}/Y_{\tau}$ 
and those in Figure~\ref{4}, show contour lines for $r_{t}=Y_{t}/Y_{\tau}$ 
in the best fits over the $(m_F,M_{1/2})$ plane.  
These parameters allow the deviation of the top, 
bottom and tau Yukawa couplings away from unification($r_{b}=r_{t}=1$). Both 
parameters show significant dependence upon $m_F$ and weak dependence upon 
$M_{1/2}$, with increasing $r_{t,\,b}$ values as we move towards 
larger $m_F$. The plots show that the level of deviation from Yukawa 
unification required for a good $\chi^2$ fit to be obtained is of the order 
of 20-35\% in $r_b$ and 0-10\% in $r_t$.  It is possible to 
account for this level of deviation through the presence of subleading 
operators, of the type mentioned in Eq.~\ref{OO'O''}, in the $33$ element 
of the Yukawa matrices. Hence the $33$ element in Eq.~\ref{SRNDmRLopMtr} 
should read,
\beq
Y_{33}=a_{33}+{\cal O} + {\cal O}^{\prime} + \ldots
\eeq
where the operators ${\cal O}$ and ${\cal O}^{\prime}$ are 
responsible for generating $r_{t,\,b}\neq 1$. The $23$ block of the neutrino 
Yukawa matrix has already shown us that a contribution to the Yukawa matrices 
from a subleading operator can actually be comparable to those from a leading 
operator. This occurs in the $23$ element of the neutrino Yukawa matrix, where
there is a contribution from the operator ${\cal O}^I$ and the 
neutrino matrix is the only one that receives a non-zero Clebsch as can be 
seen in Table~\ref{GUT}. This leads to the relative sizes of the elements
$Y_{\nu\,23}\sim 0.44$ and $Y_{\nu\,33}\sim 1$. A similar subleading 
contribution to the $33$ element of the up and down quark Yukawa matrices 
could easily account for a deviation from third family Yukawa unification 
at the level discovered in our study.

\begin{figure}[p]
\begin{center}
\scalebox{0.8}{\includegraphics*{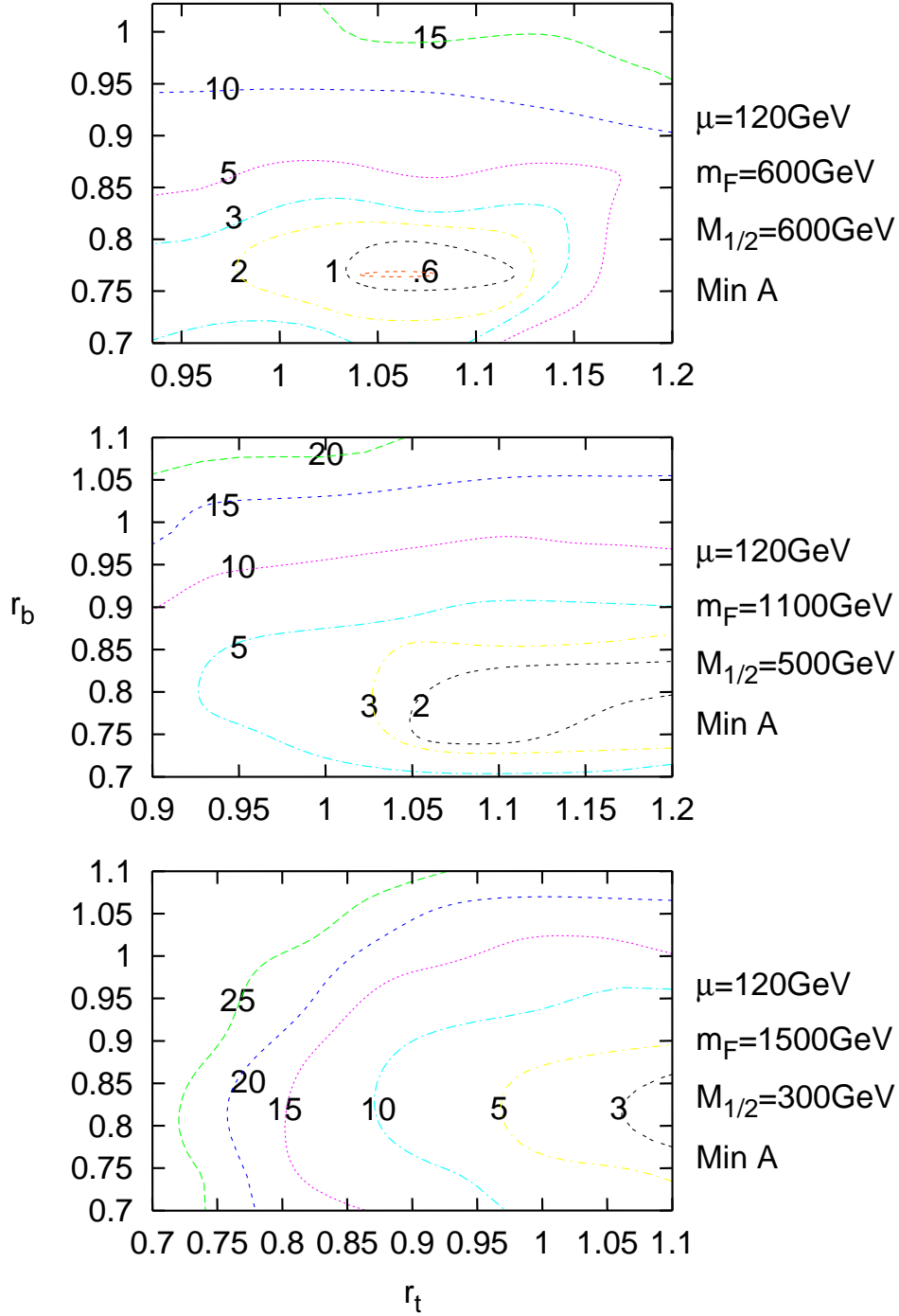}}
\vskip-20mm
\begin{minipage}[t]{15cm}
\caption{\small{$\chi^2$ contours in the $r_{t}-r_{b}$ plane.  The plots are generated with $\mu=120$~GeV and for minimum A. The three plots are each generated with fixed $M_{1/2}$, $m_F$ as labelled. The plots display the $\chi^2$ penalty which is required for exact Yukawa unification to be achieved.}}\label{29}
\end{minipage}
\end{center}
\end{figure}

Here we do not study the region in the parameter space $\;m_F>2\,$TeV, 
$A_0\approx -2\,m_F$ where the exact unification might work 
\cite{exact.yuk.unif}. Instead, we carried out a study of the additional 
$\chi^2$ penalty incurred due to demanding exact Yukawa unification in the
region  $\;m_F<2\,$TeV and $A_0=0$.  Figure~\ref{29} shows the result as 
$\chi^2$ contour plots in the $r_{t}-r_{b}$ plane corresponding to the
best fits.  The three panels were obtained from three points in the 
$(m_F,M_{1/2})$ plane and show that a very heavy penalty $\delta\chi^2 > 10$ 
is paid when requiring exact Yukawa unification in this SUSY region.

\subsection{Future Higgs searches}\label{Higgsmass}

\begin{figure}[p]
\begin{center}
\scalebox{0.8}{\includegraphics*{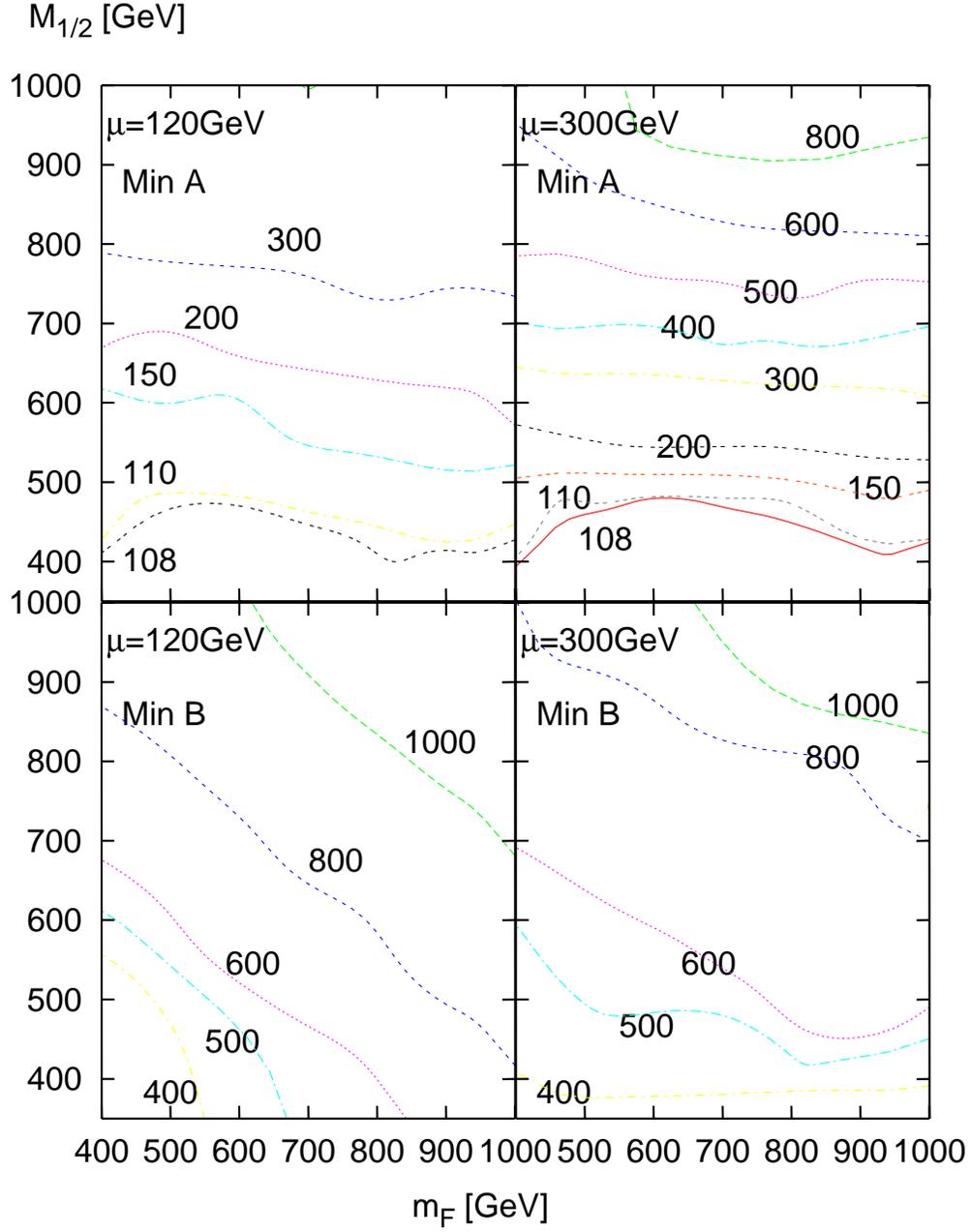}}
\vskip-20mm
\begin{minipage}[t]{15cm}
\caption{\small{Contours of the CP odd Pseudoscalar Higgs mass, $m_{A^{0}}$, are plotted in the plane of $(m_F,M_{1/2})$.  The contours are in the units of~GeV.  The four plots, are obtained from the two minima, minimum A and minimum B with $\mu=120$ and $300$~GeV as labelled.  All points in the top left corner with approximately $M_{1/2}> 700$~GeV and $m_F> 700$~GeV are unphysical due to the lightest stau becoming the LSP.}}\label{6}
\end{minipage}

\end{center}
\end{figure}

Figure~{\ref{6}} shows mass contours of the CP odd pseudoscalar Higgs, $m_{A^{0}}$. 
These plots show that for the Pseudoscalar Higgs mass minimum A prefers values approximately 
$200-300$~GeV lower than minimum B. In Figure~\ref{6} we see that for both $\mu=120$ 
and $300$~GeV, minimum A gives a very light pseudoscalar Higgs mass, $m_{A^{0}} \sim 108$~GeV, 
in the low $M_{1/2},\,m_F$ region. This is in fact the same region in which minimum A 
provides its lowest $\chi^2$. In fact Table~\ref{output} shows that for the best fit
point in minimum A we have a pseudoscalar mass of $102$~GeV and a light CP even mass of $106$~GeV.
With the Tevatron now taking data there is a high 
probability that the present lower bound on Higgs masses will be pushed higher.  
Hence we have undertaken a study of the effect this would have on our best fits.  
The plot in Figure~{\ref{32b}} shows the increase in $\chi^2$, for four points 
in the $(m_F,M_{1/2})$ plane of minimum A, due to an increase in the lower bound 
on the Higgs masses $m_{A^{0}}$, $m_{h^{0}}$.  It clearly shows that all four of 
the points can accommodate an increase in the lower bound up to approximately 
$120$~GeV, above this the $\chi^2$ increases sharply due the inability to 
accommodate such a large lower bound. 

\begin{figure}[p]
\scalebox{0.65}{\rotatebox{-90}{\includegraphics*{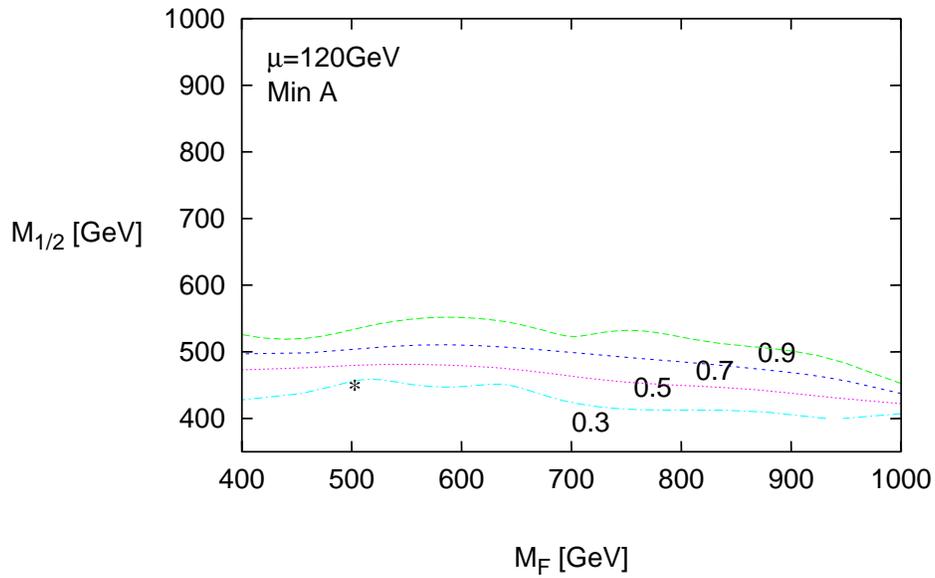}}}
\vskip20mm
\begin{center}
\begin{minipage}[t]{15cm}
\caption{\small{Contours of $\sin(\beta - \alpha)$, which defines the strength of the $Z$ boson coupling to the Higgs $h^{0}$ relative to that of $H^{0}$. For values of $\sin(\beta-\alpha)$  near one the $Z-h^{0}$ coupling is large and for small values the $Z-H^{0}$ coupling is large. The contours are plotted using data from minimum A with $\mu=120$~GeV. The the best fit point at $M_{1/2}=450$~GeV, $m_{f}=500$~GeV is marked with an asterisk.}}\label{sinb-a}
\end{minipage}

\end{center}
\end{figure}

The coupling of the light CP even Higgs, $h^{0}$, to the $Z$ boson is proportional to 
$\sin(\beta-\alpha)$ and that of the heavy CP even Higgs, $H^{0}$, is proportional to
$\cos(\beta-\alpha)$, where $\alpha$ is the mixing angle for the CP even Higgs states. 
In figure~\ref{sinb-a}, which shows contours of $\sin(\beta-\alpha)$ for
points in minimum A,
we see that in the low $M_{1/2}$ region $\sin(\beta-\alpha)$ is small and hence 
% we have the unusual situation of 
the $Z$ couples dominantly to the heavier Higgs state $H^{0}$, rather that the lighter $h^{0}$.  
Therefore, in this region it is the heavier state, $H^{0}$, 
which is the standard model like Higgs and so the LEP limit will apply to the larger $m_{H^{0}}$ 
and not $m_{h^{0}}$.
Table~\ref{output} shows that we have exactly this situation for the best fit points of minimum A
where $\sin(\beta-\alpha)\sim 0.2$, therefore the standard model like Higgs is the heavier state 
$H^{0}$ for these points with a mass of $113$~GeV. Assuming a $3$ GeV error in our numerical 
calculation means we are compatible with the present LEP limit of $114.4$ GeV.

We have checked that the light Higgs spectrum is consistent with the current limit on the 
rate for $B_s\to\mu^+\mu^-$. \cite{Bs.mumu}

\subsection{CHOOZ angle, $\theta_{13}$}

\begin{figure}[p]
\begin{center}
\scalebox{0.7}{\includegraphics*{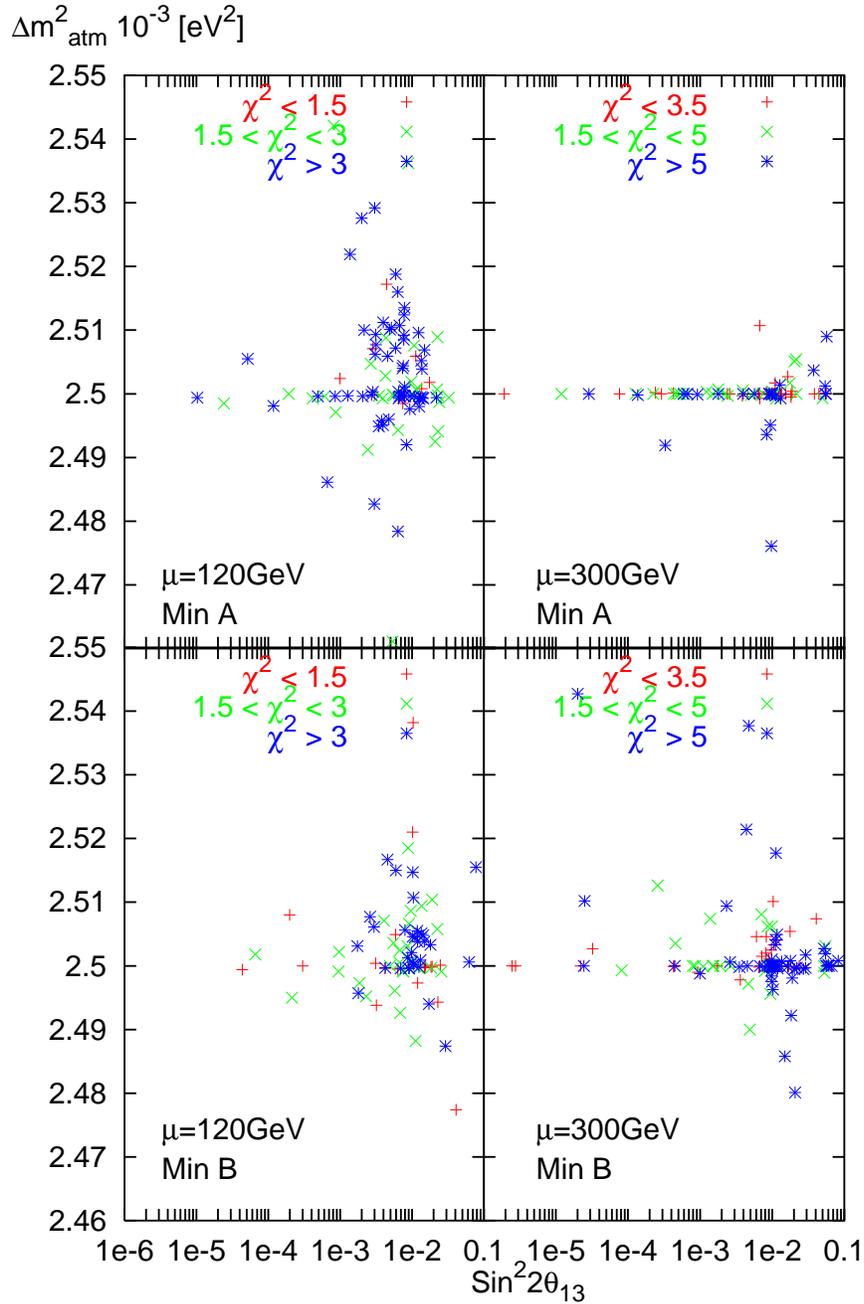}}
%\vskip-20mm
\begin{minipage}[t]{15cm}
\caption{\small{The four panels contain scatter plots of the values of $\Delta m^2_{Atm}$ against $\sin^2 2 \theta_{13}$ coming from the best fit points in the $(m_F,M_{1/2})$ plane. Each plot shows results obtained from either minimum A or minimum B, with $\mu=120$ or $300$~GeV as labelled.  The points are grouped according to their $\chi^2$ values as inticated.}}\label{14}
\end{minipage}

\end{center}
\end{figure}

Figure~\ref{14} shows scatter plots of $\sin^2 2\theta_{13}$ against $\Delta m^2_{Atm}$ 
for both minimum A and minimum B with $\mu = 120$ and $300$~GeV.  Each point denotes 
values obtained from individual points in the $(m_F,M_{1/2})$ plane, with points 
grouped according to the value of $\chi^2$.  These plots show that the Model can 
easily yield values of $\theta_{13}$ that are within the present CHOOZ limit, 
$\theta_{13}\lesssim 0.22$.  Each of the plots in Figure~\ref{14} shows that the 
best fit points, denoted by a $+$ symbol, give a range of values of $\sin^2 2\theta_{13}$ 
from, $10^{-6}$ to $0.1$. Although our results do not give any firm prediction for the 
value of $\theta_{13}$, it can be seen that the model favours the region, 
$10^{-4}<\sin \theta_{13}<0.1$, just below the present CHOOZ limit.

\section{Summary and Conclusion\label{conc}}

\begin{figure}[p]
\begin{center}
\scalebox{0.8}{\includegraphics*{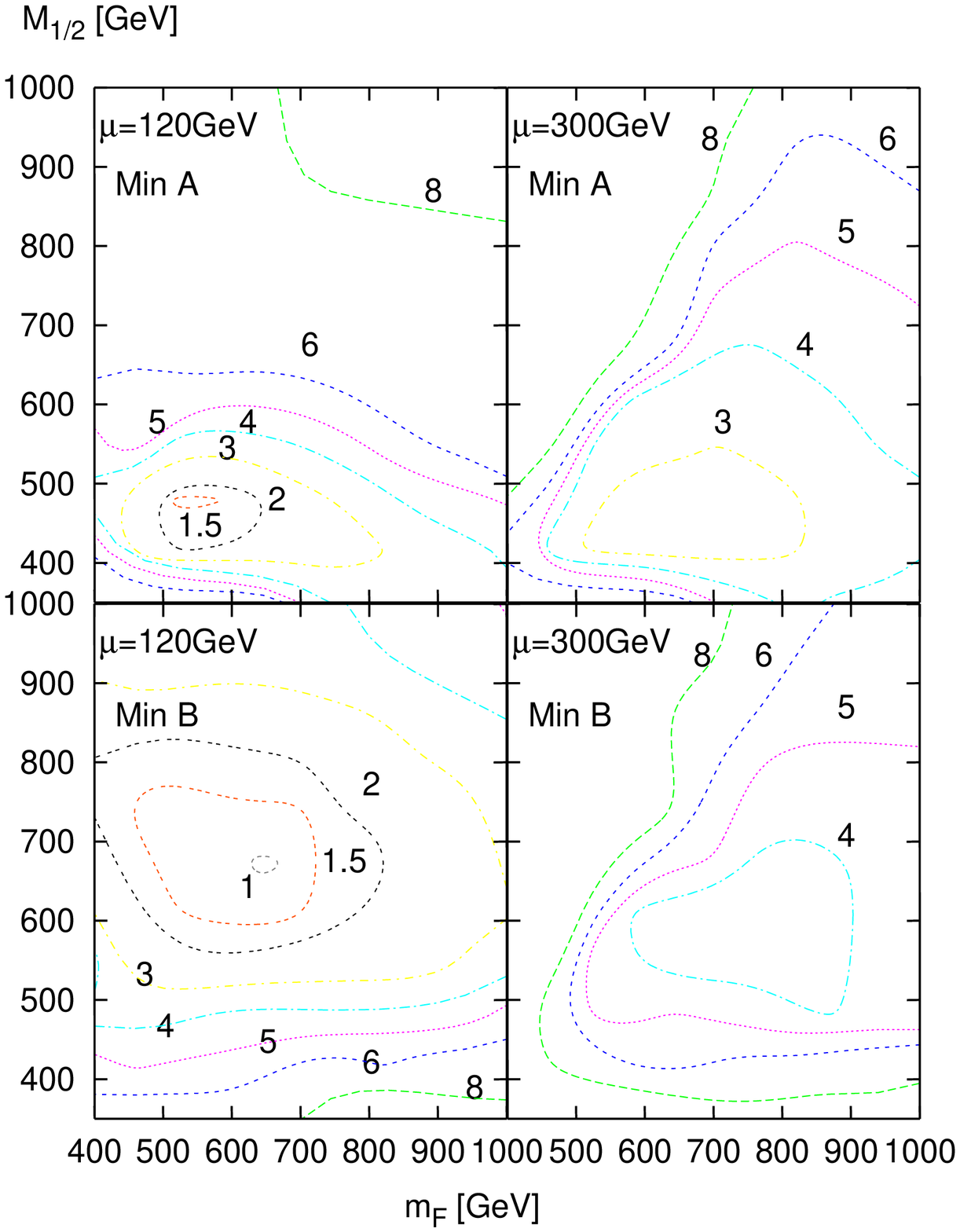}}
\vskip-20mm
\begin{minipage}[t]{15cm}
\caption{\small{$\chi^2$ contour plot in the plane of $(m_F,M_{1/2})$.  The four plots, are obtained from the two minima, minimum A and minimum B with $\mu=120$ and $300$~GeV as labelled.  All points in the top left corner with approximately $M_{1/2}> 700$~GeV and $m_F> 700$~GeV are unphysical due to the lightest stau becoming the LSP.}}\label{1}
\end{minipage}

\end{center}
\end{figure}

We have performed a complete global phenomenological analysis of
a realistic string-inspired model
based on the supersymmetric Pati-Salam $SU(4)\times SU(2)_L \times SU(2)_R$ 
 gauge group supplemented by a $U(1)$ family symmetry.
Global contour plots in the 
$(m_F,\,M_{1/2})$ plane have been presented in Figure~\ref{1}, showing two 
$\chi^2$ minima.
 These two distinct minima differ numerically 
by the relative sign of the D-term.  This gives interesting 
phenomenological differences between the two minima, notably one has 
$BR(\tau \rightarrow \mu \gamma)$ near the present limit and a 
heavy pseudoscalar Higgs $m_{A^{0}}$, while the other has 
$BR(\tau \rightarrow \mu \gamma)$ well below the present bound but 
a light pseudoscalar Higgs $m_{A^{0}}$. Both minima 
give a good fit to the present muon $g-2$ discrepancy over a large 
region of parameter space and give $\sin^2 2\theta_{13}$ over the 
range $10^{-5}-0.1$. Our best fit predictions for the
superpartner masses for each of the two minima for two different
$\mu$ values are summarised in Table~\ref{output}. 

We emphasise again that our analysis really should be considered
as consisting of two distinct parts, associated with flavour physics on the one
hand and unification and electroweak symmetry breaking on the other hand.
For the flavour part, we have proposed a complete model 
in Table~\ref{GUT} which gives an accurate description of
all fermion masses and mixing angles,
including the LMA MSW neutrino solution.
We have shown that improved limits on
$BR(\tau \rightarrow \mu \gamma)$
could begin to rule out one of our two minima.
The conclusions on $BR(\tau \rightarrow \mu \gamma)$ are applicable 
to a wide class of models which achieve approximate maximal 
atmospheric neutrino mixing via the see-saw mechanism in the MSSM with 
a large 23 entry in the neutrino Yukawa matrix.
On the other hand $BR(\mu \rightarrow e \gamma)$ is predicted to be about 
two orders of magnitude below the current limit, which is a
consequence of the specific flavour structure of the model in
Table~\ref{GUT}. 

Regarding unification, 
the model predicts approximate third family Yukawa unification and hence 
large $\tan \beta \sim 50$. Electroweak symmetry breaking was achieved
with the help of D-terms and non-universal soft Higgs mass,
which allows small $\mu$ values.
The property of exact Yukawa unification was 
relaxed throughout the analysis and it was found that 
a deviation of 20-35\% for the bottom Yukawa coupling 
and 0-10\% for the top Yukawa coupling are required 
for a good fit to be obtained. We showed that 
relaxing Yukawa unification has the
effect of allowing small values of the soft scalar mass $m_F$,
and lighter squark and slepton masses as a consequence.

\begin{figure}[p]
\begin{center}
\rotatebox{-90}{\scalebox{0.5}{\includegraphics*{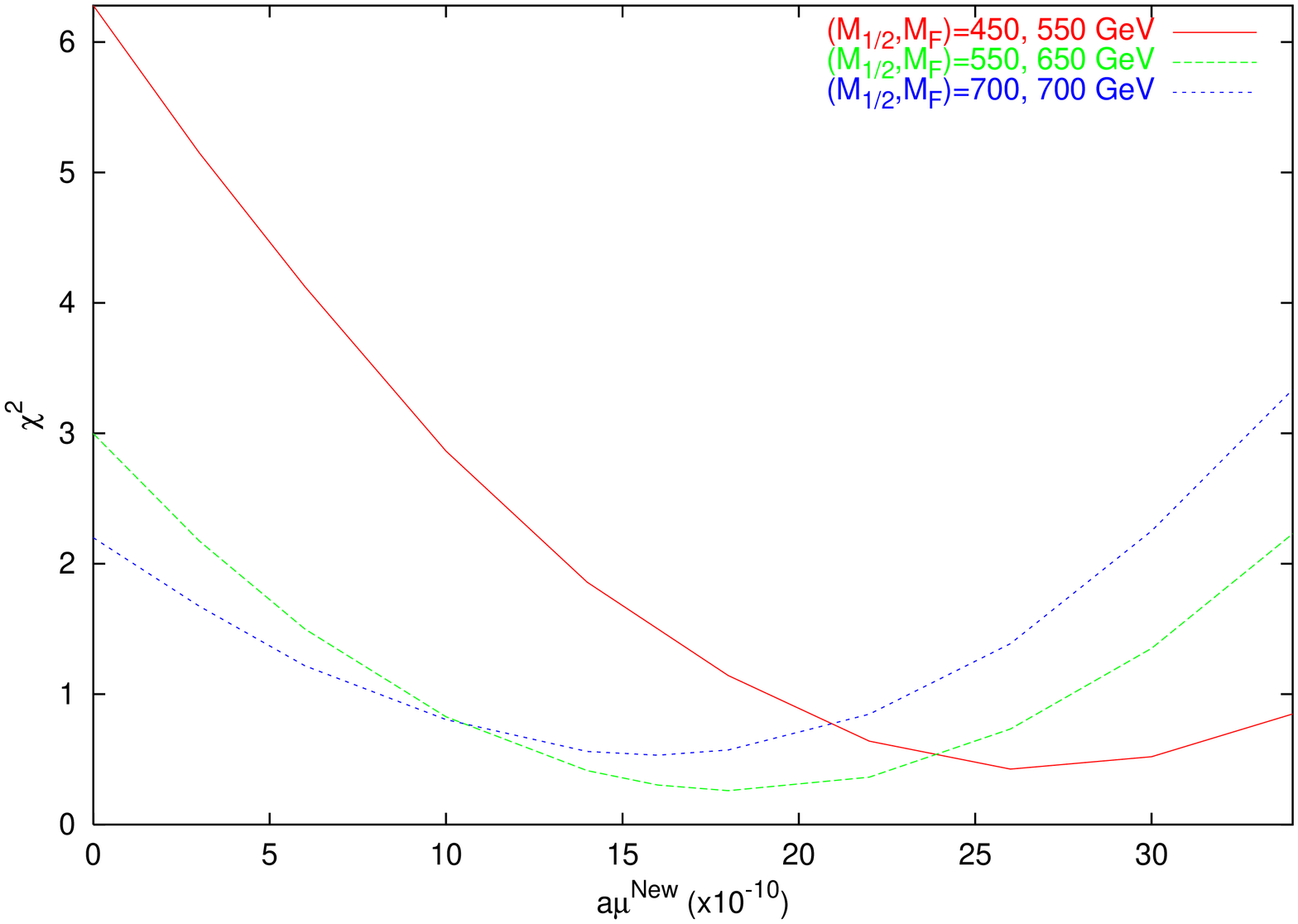}}}
\begin{minipage}[t]{15cm}
\caption{\small{This plot displays the effect on $\chi^2$ due to a future change in the value of the muon $g-2$ discrepancy. The value of the muon anomalous magnetic moment is varied from the present value of $34\times 10^{-10}$ down to zero. The resulting change in $\chi^2$ is observed for three points in the $(m_F,M_{1/2})$ plane.}}\label{32a}
\end{minipage}
%\vskip10mm
\rotatebox{-90}{\scalebox{0.5}{\includegraphics*{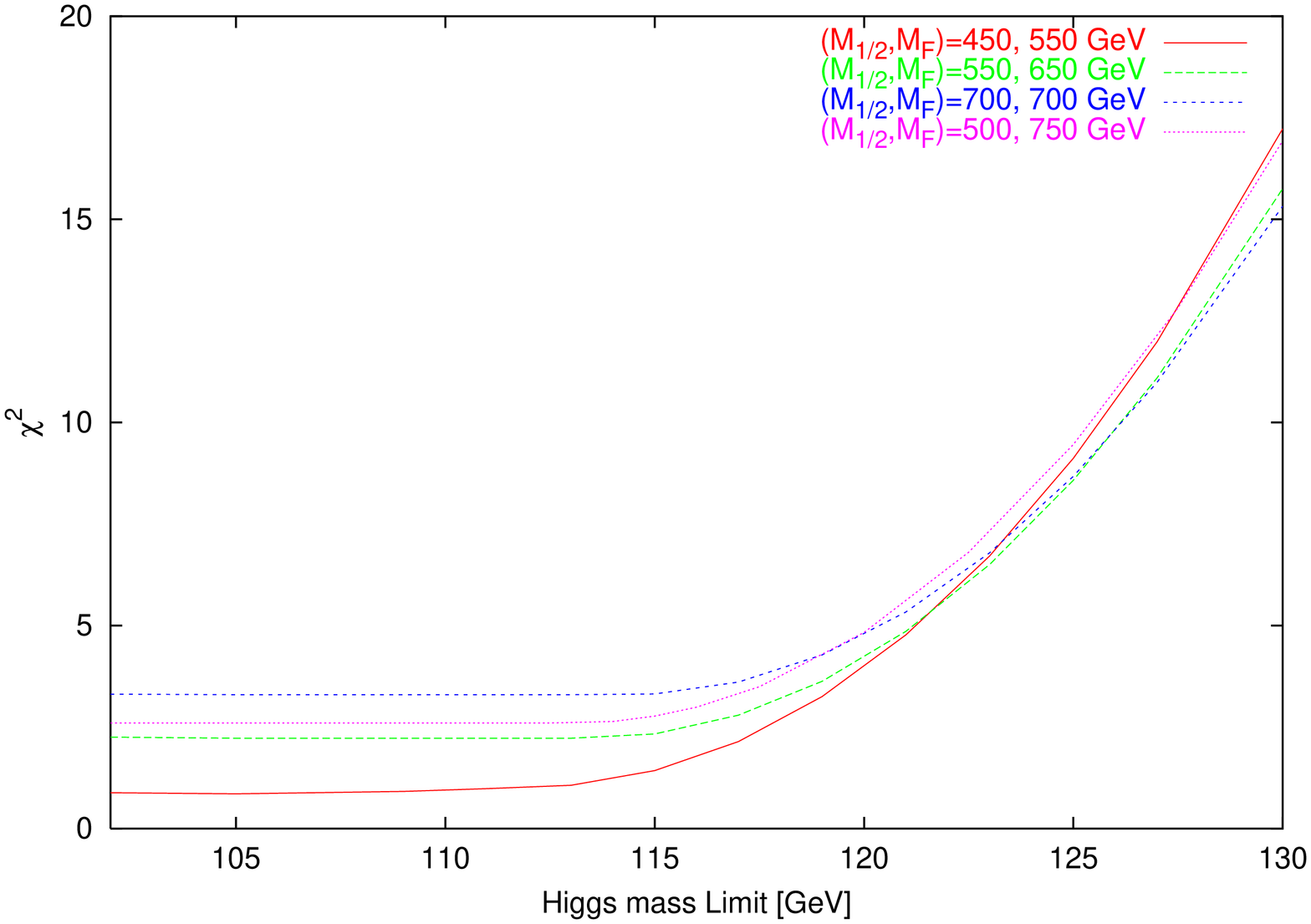}}}

%\vskip-10mm
\begin{minipage}[t]{15cm}
\caption{\small{This plot displays the effect on $\chi^2$ due to an increase in the lower bound on the Higgs mass from direct searches. As in Figure~\ref{32a} the variation in $\chi^2$ is observed through individual points in the $(m_F,M_{1/2})$ plane.}}\label{32b}
\end{minipage}

\end{center}
\end{figure}

Further studies of the effects of future direct Higgs searches and a change 
to the present muon g-2 discrepancy are shown in 
Figures~\ref{32a} and \ref{32b}.
We found that our best 
fit points, for the minima with lighter Higgs masses, can accommodate 
a lower bound on Higgs masses up to about $120$~GeV. For these points the coupling of the 
lighter CP even Higgs state to the $Z$ boson is suppressed, leaving
the heavier of the two CP even states acting as the standard model like Higgs.

In conclusion, we have constructed and analysed a complete supersymmetric
Pati-Salam model which agrees with all laboratory observables
and constraints. Using a global analysis we identify the most
preferred regions of the SUSY parameter space, and find
a rather light superpartner spectrum corresponding to 
$(m_F,M_{1/2})\sim (600,600)$ (in GeV) well within reach
of the LHC.

\vskip 0.1in
\noindent
 {\large {\bf Acknowledgments}}\\
S.K. thanks PPARC for a Senior Fellowship and J.P. thanks PPARC for a studentship.
\newpage

\newpage
\clearpage

%\section{Global Analysis Results with $\mu = 120$~GeV}\label{B}
%{Numerical Results with $\mu = 120$~GeV}

%%%%%%%%%%%%%
\newpage
\clearpage

%%%%%%%%%%%%%
\newpage
\clearpage

%%%%%%%%%%%%%
\newpage
\clearpage

%%%%%%%%%%%%%
\newpage
\clearpage

%%%%%%%%%%%%%
\newpage
\clearpage

%%%%%%%%%%%%%
%\newpage
%
%\begin{figure}[p]
%\begin{center}
%\scalebox{0.8}{\includegraphics*{bsg.ps}}
%\vskip-20mm
%\begin{minipage}[t]{15cm}
%\caption{\small{BR$(b \rightarrow s \gamma)$ contour plot in the plane of $(m_F,M_{1/2})$.  The four plots, are obtained from the two minima, minimum A and minimum B with $\mu=120$ and $300$~GeV as labelled.  All points in the top left corner with approximately $M_{1/2}> 700$~GeV and $m_F> 700$~GeV are unphysical due to the lightest stau becoming the LSP. The above plots show contours for the present experimental value, $3.47\times 10^{-10}$, and $1$ and $2$ $\sigma$ regions.}}\label{2b}
%\end{minipage}
%
%\end{center}
%\end{figure}
%%%%%%%%%%%%%
\newpage
\clearpage

%%%%%%%%%%%%%
\newpage

%%%%%%%%%%%%%
\newpage
\clearpage

%\section{Deviation from Yukawa Unification}\label{D}

%%%%%%%%%%%%%
\newpage
\clearpage

\newpage
\clearpage

%\section{Study of a future measurement of BR$(\tau\rightarrow \mu \gamma)$}\label{E}
%{Study of a future measurement of BR$(\tau\rightarrow \mu \gamma)$}

%\begin{figure}[p]
%\begin{center}
%\scalebox{0.7}{\includegraphics*{tmg_var.ps}}
%\vskip-20mm
%\begin{minipage}[t]{15cm}
%\caption{\small{These plots show data taken with $M_{1/2}=m_F=550$~GeV and $\mu=120$~GeV.  The data was produced by introducing an imaginary measurement of $\tau \rightarrow \mu \gamma$ and gradually lowering its value.  We started with BR$(\tau \rightarrow \mu\gamma)=2\times 10^{-6}$, with a point in minimum B and then lowered the branching ratio down to $0.05\times 10^{-6}$.  The upper plot shows the variation of total $\chi^2$ as the imaginary measurement is decreased.  At the same time the lower plot shows the variation of $m_{D}=|D^2|$, the magnitude of the broken D-term.}}\label{31}
%\end{minipage}
%
%\end{center}
%\end{figure}

%%%%%%%%%%%%%
%\newpage
%
%\begin{figure}[p]
%\begin{center}
%\rotatebox{-90}{\scalebox{0.4}{\includegraphics*{tmg_var_chi.ps}}}
%\vskip5mm
%\rotatebox{-90}{\scalebox{0.4}{\includegraphics*{tmg_var_m_D.ps}}}
%
%\vskip-10mm\begin{minipage}[t]{15cm}
%\caption{\small{Plot}}\label{32}
%\end{minipage}
%
%\end{center}
%\end{figure}
%%%%%%%%%%%%%%%%%%
\newpage

\end{document}